## CONTENTS

# Generative AI as Digital Media

Gilad Abiri*

### Abstract

The hype surrounding Generative AI paints it as revolutionary and potentially apocalyptic and calls for equally novel regulation. This essay argues that such an approach is misguided. It shows that generative AI is best understood as the next step in the evolution, rather than a revolution, of our algorithmic media landscape, following in the footsteps of search engines and social media. Together, these digital media platforms centralize information control, use complex algorithms to shape content, and rely heavily on data. These platforms also create shared problems: unchecked power, echo chambers, and the erosion of traditional gatekeepers.

It follows that we should approach their regulation with the same goal: Media institutions must be trusted and trustworthy. Without this trust, public discourse risks devolving into isolated echo chambers where only comforting, tribally-approved beliefs survive—a threat exacerbated by generative AI's ability to bypass gatekeepers and tailor "truth." Regulation must foster accountability, transparency, and environments that inspire public confidence towards generative AI platforms.

Risk regulation, the dominant approach in current AI governance, emphasizes reactive risk mitigation. Both the European Union's AI Act and the United States' Executive Order 14110 on Ensuring Trustworthy AI prioritize identifying and mitigating measurable risks. This approach excels at preventing crises in areas like national security, public health, and algorithmic bias. It is a good way of dealing with AI as a revolutionary, unpredictable, new technology. However, this Article shows that its focus on measurable risk makes it ill-suited to address the dimensions of building trust in digital media platforms.

---

* Assistant Professor of Law, Peking University School of Transnational Law and Visiting Fellow, Information Society Project, Yale Law School.

*Achieving this demands not just risk reduction, but proactive, public-oriented measures.*

*If we continue to understand generative AI as a completely revolutionary technology necessitating reactive regulation, we risk repeating past mistakes that left social media and search engines unregulated for decades. We must ask how to proactively shape an algorithmic media landscape serving the public good—one that cultivates quality information and civil discourse.*

## Introduction

The way we imagine a new technology plays a pivotal role in how we regulate it. We are told by big tech that the introduction of Generative AI ('GenAI') "is more important than fire or electricity."[1] That it "has the potential to revolutionize nearly every industry."[2] That it is "more dangerous than nukes."[3] The theatrical nature of these proclamations, and the aura of mystique surrounding the term `AI', inevitably impacts the way we conceptualize and implement its regulation. We are led to believe GenAI represents a dramatic breakthrough necessitating equally revolutionary regulations, befitting a technology with monumental, frightening, and uncertain impacts. Yet this narrative promoted by the developers of GenAI is misleading.

GenAI, particularly in the realm of digital media, signifies more of an evolution than a revolution. It represents a continuation of trends that have long been in motion. Over the last two decades, two types of algorithms have become central in shaping public discourse: those that curate the content we encounter on digital platforms,[4] and those that govern the moderation of the content users contribute.[5] Presently, we are witnessing the emergence of a third kind of algorithm: one specialized in generating human-like content.⁶ More tangibly, over the past two decades our media ecosystem has been overtaken by search engines and social media platforms, which are now being joined by algorithmic chatbots. These are all technologies driven by complex machine learning programs that facilitate media consumption. In this Article, I argue against viewing generative algorithms as some new epoch-making technology. Rather, we should see them for what they are—the next phase in the steady progression of algorithmic mediation over our information. GenAI continues the trajectory of search engines and social platforms in algorithmizing content. Therefore, regulating GenAI is fundamentally linked to regulating other algorithmic systems governing media and knowledge.

---

[1] Catherine Clifford, *Google CEO: A.I. is More Important Than Fire or Electricity*, CNBC (Feb. 1, 2018, 12:56 PM), https://www.cnbc.com/2018/02/01/google-ceo-sundar-pichai-ai-is-more-important-than-fire-electricity.html/ [https://perma.cc/E6C7-M3PB].

[2] Samantha Kelly, *Sam Altman Warns AI Could Kill Us All. But He Still Wants the World to Use It*, CNN (Oct. 31, 2023, 6:00 AM), https://edition.cnn.com/2023/10/31/tech/sam-altman-ai-risk-taker/index.html/ [https://perma.cc/E8G6-MM2E].

[3] Catherine Clifford, *Elon Musk: 'Mark My Words — A.I. is Far More Dangerous Than Nukes'*, CNBC (Mar. 14, 2018, 11:31 AM), https://www.cnbc.com/2018/03/13/elon-musk-at-sxsw-a-i-is-more-dangerous-than-nuclear-weapons.html/ [https://perma.cc/R6MQ-SP8Y].

[4] Gilad Abiri & Xinyu Huang, *The People's (Republic) Algorithms*, 12 Notre Dame J. Int'l & Comp. L. 16, 19-20 (2022).

[5] Gilad Abiri, *Moderating from Nowhere*, 47 BYU L. Rev. 757, 772 (2022).

My argument proceeds in four stages:

In Part I, I put forth the idea of grouping social media, search engines, and generative algorithms together under the single concept of *digital media platforms*. I argue these should be seen as interconnected technologies that warrant a unified regulatory approach. It makes sense to consider digital media platforms together since they both share fundamental qualities and raise similar societal concerns. They are defined by their algorithmic backbone for key functions like content filtering, recommendation engines, and generating novel content.⁷ This algorithmic foundation is intrinsically tied to their data-driven nature, where accumulating and analyzing vast datasets is imperative for refining and personalizing user experiences.⁸ Lastly, their global reach and concentration of control within a few dominant entities signify a major shift from being a myriad of mostly local, media organizations to being a public sphere dominated by 3–4 major global technology corporations.[9]

---

⁶ *See, e.g.*, Harshvardhan GM et al., *A Comprehensive Survey and Analysis of Generative Models in Machine Learning*, 38 Comput. Sci. Rev. 1 (2020); Keng-Boon Ooi et al., *The Potential of Generative Artificial Intelligence Across Disciplines: Perspectives and Future Directions*, J. Comput. Info. Sys. 1 (2023); Francesca Grisoni et al., *Combining Generative Artificial Intelligence and On-Chip Synthesis for De Novo Drug Design*, 7 Sci. Adv. 1 (2021); Zhuoxuan Jiang et al., *Leveraging Key Information Modeling to Improve Less-Data Constrained News Headline Generation via Duality Fine-Tuning*, 1 Proc. 2ⁿᵈ Conf. Asia-Pacific Chapter Ass'n for Computational Linguistics & 12ᵗʰ Int'l J. Conf. on Nat. Language Processing 57 (2022); Simon Zhai et al., *Enabling Predictive Maintenance Integrated Production Scheduling by Operation-Specific Health Prognostics with Generative Deep Learning*, 61 J. Mfg. Sys. 830 (2021); David Baidoo-Anu & Leticia O. Ansah, *Education in the Era of Generative Artificial Intelligence (AI): Understanding the Potential Benefits of ChatGPT in Promoting Teaching and Learning*, 7 J. A.I. 52 (2023); Steven J. Quan, James Park & Sugie Lee, *Artificial Intelligence-Aided Design: Smart Design For Sustainable City Development*, 46 Env't & Plan. B: Urb. Analytics & City Sci. 1581, 1584 (2019).

⁷ *See* discussion *infra* Part I.B.

⁸ *Id.*

Since they share fundamental characteristics, the challenges arising from social media, search engines, and generative AI are closely interconnected. First, the centralization of power in these platforms exacerbates problems around information control, privacy, and potential for abuse.[10] Second, the reliance on algorithms to curate and recommend content has created echo chambers, where users are increasingly exposed to information that affirms their existing views, diminishing viewpoint diversity and undermining democratic exchange of ideas.[11] Third, these platforms contribute to the bypass effect, where traditional local gatekeepers and norms are sidelined in favor of algorithmic content dissemination, challenging regulatory frameworks and cultural contexts that have historically governed speech and information flow.[12] Ultimately, the trend of personalization of GenAI is likely to lead to narrower and narrower echo chambers, a more polarizing side effect than that of social media, isolating individuals from the public forum.

Having established a shared focal point for regulation, Part II turns to the goals of regulating digital media platforms, including Generative AI. The question arises: If Generative AI is a new type of digital intermediary, what should we aim to achieve by regulating it?

The primacy of digital media platforms has transformed global public spheres. While we once celebrated this shift,[13] it is now implicated in perpetuating social problems like the spread of hate speech and misinformation.[14] Scholars like Jack Balkin highlight a critical shortcoming in our digital era: the lack of "trusted and trustworthy intermediaries"[15] to facilitate, organize, and curate public discourse.[16] This deficiency jeopardizes any public sphere, as without trust in institutions responsible for delineating reliable knowledge and acceptable speech, society risks devolving into a rhetorical battlefield marked by tribalism and comfortable beliefs, undermining foundational free speech values.[17]

---

[9] *Id.*

[10] *See, e.g*, Lina M. Khan, *The Separation of Platforms and Commerce*, 119 Colum. L. Rev. 973 (2019) (describing how the size of big tech creates myriad social harms); Amy Kapczynski, *The Law of Informational Capitalism*, 129 Yale L. J. 1460 (2020) (exploring the implications of the power of big tech and their reliance on information for profit); Juho Lindman, Jukka Makinen & Eero Kasanen, *Big Tech's Power, Political Corporate Social Responsibility and Regulation*, 38 J. Info. Tech. 144, 145, 152 (2023).

[11] *See* Gilad Abiri & Johannes Buchheim, *Beyond True and False: Fake News and the Digital Epistemic Divide*, 29 Mich. Telecomm. & Tech. L. Rev. 59 (2022) (describing the rise of digital epistemic divide); *see also* discussion *infra* Part I.B.2.

[12] *See generally* Axel Bruns, Gatewatching: Collaborative Online News Production 11 (2005) (describing the new phenomenon of gate watching).

[13] *See* Yochai Benkler, Hal Roberts, Robert Faris, & Alicia Solow Nierderman, *Social Mobilization and the Networked Public Sphere: Mapping the SOPA-PIPA Debate*, 32 Pol. Commc'n 594 (2015) (supporting an optimistic view of the potential of tech media for networked democratic participation); *see also* Yochai Benkler, *A Free Irresponsible Press: Wikileaks and the Battle Over the Soul Of the Networked Fourth Estate*, 46 Harv. C.R.–C.L. L. Rev. 311, 311 (2011) (using WikiLeaks as an example to show how the Internet enables individuals to speak their mind).

The path towards establishing digital media platforms, including Generative AI, as trusted and trustworthy intermediary institutions is impeded by two key trust deficits: The first centers on misaligned incentives, where the economic models driving these platforms often prioritize engagement and revenue over public welfare.[18] This misalignment fosters environments where misinformation and sensationalism thrive at the expense of societal well-being. The second deficit stems from an unfamiliarity gap arising from the global nature of these platforms, distancing them from users' localized contexts. This gap is marked by a lack of deep cultural and community integration, making it difficult for platforms to engender trust and belonging.[19]

Having laid out the target and aim of GenAI regulation, Part III illustrates the inadequacy of the prevailing AI regulation approach in achieving this objective. Current attempts at AI regulation, typified by the European Union's Artificial Intelligence Act (AI Act)[20] and the United States' Executive

---

[14] *See* Gilad Abiri & Sebastian Guidi, *From a Network to a Dilemma: The Legitimacy of Social Media*, 26 Stan. Tech. L. Rev. 92, 139 (2023) [hereinafter Abiri & Guidi, *From a Network to a Dilemma*].

[15] Jack M. Balkin, *How to Regulate (and Not Regulate) Social Media*, 1 J. Free Speech L. 71, 79 (2021) [hereinafter Balkin, *To Regulate*].

[16] *See id.; see also* Jack M. Balkin, *To Reform Social Media, Reform Informational Capitalism*, *in* Social Media, Freedom of Speech, and the Future of our Democracy 234 (Lee C. Bollinger & Geoffrey R. Stone eds., 2022) [hereinafter Balkin, *To Reform*].

[17] *See* Balkin, *To Regulate*, *supra* note 15, at 79 ("Without these trusted institutions and professions, the practices of free expression become a rhetorical war of all against all."); *see also* Balkin, *To Reform*, *supra* note 16, at 242 ("Weaken the institutions or destroy trust, and the public sphere becomes a rhetorical war of all against all, where no one is believed except the members of one's own tribe, and people cleave to whatever beliefs are most comforting to them").

[18] *See* discussion *infra* Part II.A.

[19] *See* discussion *infra* Part II.B.

[20] *Proposal for a Regulation of the European Parliament and of the Council Laying down Harmonised Rules on Artificial Intelligence (Artificial Intelligence Act) and Amending Certain Union Legislative Acts*, at 25, COM (2021) 206 final (Apr. 21, 2021) [hereinafter *AI Act*].

Order 14110 on Ensuring Trustworthy AI (hereinafter Executive Order 14110),[21] concentrate on risk management and mitigation. These regulatory frameworks prioritize identifying and mitigating risks associated with AI technologies, categorizing AI systems based on their potential harm to ensure safety and compliance. However, while these measures are crucial, they inadvertently overlook the integral role of GenAI as a media entity, central to public discourse and societal narrative shaping.

For instance, the AI Act's broad categorization of AI systems into unacceptable, high, or low/minimal risk groups,[22] and Executive Order 14110's focus on regulating high-risk foundation models, illustrate the centrality of risk management.[23] These frameworks aim to safeguard against tangible harms, such as privacy violations or discriminatory outcomes.[24] Yet, they do not fully grapple with the subtler, yet equally significant, impact of GenAI on the digital public sphere—such as the dissemination of information, the formation of public opinion, and the potential for echo chambers and misinformation.[25]

The inadequacy of these approaches becomes apparent when considering the trust deficits that plague digital media platforms. The misalignment of incentives and the unfamiliarity gap are not issues that can be resolved through risk mitigation strategies alone. For example, while the AI Act and Executive Order 14110 may enforce transparency and data governance, which may contribute some to the creation of trust, these measures do not directly address the economic models that drive platforms to prioritize engagement over accuracy or the global-local divide that hampers community building and trust.

Finally, Part IV builds a regulatory bridge between social media and GenAI, emphasizing how strategies developed for social platforms can be effectively applied to GenAI. This part explores regulatory tools beyond risk management, focusing on policies that seek to align digital platform incentives with user interests and mitigate the unfamiliarity between global platforms and local users.

---

[21] Exec. Order No. 14110, 88 Fed. Reg. 75191 (Oct. 30, 2023).

[22] European Parliament News 20230601STO93804, EU AI Act: First Regulation on Artificial Intelligence (Dec. 19, 2023, 11:45 AM), https://www.europarl.europa.eu/pdfs/news/expert/2023/6/story/20230601STO93804/20230601STO93804_en.pdf [https://perma.cc/3SWZ-EW3Z] [hereinafter EU AI ACT News].

[23] Marianna Drake, Marty Hansen, Lisa Peets, Will Capstick, Jayne Ponder, et al., *From Washington to Brussels: A Comparative Look at the Biden Administration's Executive Order and the EU's AI Act*, Compliance & Enforcement (Nov. 30, 2023) (describing one of the areas of commonality between the EO and the AI Act as their focus on high-risk AI); *see also* Exec. Order No. 14110, 88 Fed. Reg. 75191, 75194, 75196 (Oct. 30, 2023).

[24] *See* Exec. Order No. 14110, 88 Fed. Reg. 75191, 75192 (Oct. 30, 2023).

[25] Urbano Reviglio & Claudio Agosti, *Thinking Outside the Black-Box: The Case for "Algorithmic Sovereignty" in Social Media*, 6 Soc. Media + Soc'y 1, 1, 5 (2022) (describing several media-harms of algorithmic curation).

The discussion begins by examining potential reforms to liability shields like Section 230 of the Telecommunications Act of 1996 to better align platform operations with societal well-being, suggesting adjustments could compel GenAI platforms to minimize harmful content while preserving free speech.[26] It also considers how increased competition and imposed interoperability could incentivize prioritizing user welfare, drawing parallels to social media regulation where competition improves content moderation and user engagement.[27] Additionally, the concept of information fiduciaries is proposed as a model for GenAI, emphasizing the duty of platforms to protect user interests, particularly regarding personal data.[28] This aims to shift business models away from exploiting user information towards prioritizing user welfare and ethical data use.

To address the familiarity trust deficit, the Article highlights the importance of incorporating local community insights into the governance of GenAI platforms.[29] By engaging local civil society in content moderation and policy formation, GenAI can better reflect and respect diverse cultural norms and values, bridging the gap between global technology and local contexts. This approach aims to foster a more trusted and culturally coherent digital public sphere, leveraging lessons from social media regulation to address the unique challenges posed by GenAI.

The Article ends with a comparative analysis of the EU's AI Act and Digital Services Act (DSA).[30] This is meant to show that the legal ramifications of seeing GenAI as a part of digital media platforms are both immediate and meaningful. The analysis highlights the superior suitability of the DSA for regulating GenAI's media dimensions, given its explicit focus on online media platforms. Consequently, the DSA emerges as a more pertinent choice than the AI Act for addressing the unique challenges posed by GenAI.

---

[26] 47 U.S.C. § 230; Leslie Kendrick, *Speech, Intent, and the Chilling Effect*, 54 Wm. & Mary L. Rev. 1633, 1633 (2013) ("Imposing strict liability for harmful speech, such as defamatory statements, would overdeter, or chill, valuable speech, such as true political information.").

[27] *See* Balkin, *To Reform*, *supra* note 16, at 247 ("With more platforms vying for user attention, companies will have 'greater incentives to give end users what they want from social media' including improved content moderation policies and practices.").

[28] Jack M. Balkin, *Information Fiduciaries and the First Amendment*, 49 U.C. Davis L. Rev. 1183, 1207-08 (2016) [hereinafter Balkin, *Information Fiduciaries*].

[29] *See* discussion *infra* Part VI.B.

[30] *AI Act*, *supra* note 20; Eur. Parliament & Eur. Council, *The Digital Services Act: Ensuring a Safe and Accountable Online Environment*, https://commission.europa.eu/strategy-and-policy/priorities-2019-2024/europe-fit-digital-age/digital-services-act_en/ [https://perma.cc/7KM2-QAZ2] [hereinafter *The DSA Policy Essay*].

I. Generative AI as Digital Information Platform

This part argues that we should understand GenAI as a subset within a broader category of digital media platforms, which includes entities such as search engines and social media. It opens by briefing the reader on Generative AI, providing baseline knowledge. It then defines and advocates for the concept of digital media platforms that are distinct from traditional media institutions and from other algorithmic products, grouping GenAI with entities like search engines and social media. This classification clarifies GenAI's role and connects it to the broader digital landscape, emphasizing its relationship with other key platforms.

Section A outlines the core mechanisms of GenAI, its training process, and its capabilities in content creation, showcasing its versatility in various media forms and its applications beyond media. It also touches on the emergence of Large Language Models (LLMs) and their ability to integrate current information, highlighting the shift they represent in the digital information ecosystem.

Section B discusses GenAI in the context of digital media platforms, examining its role in the bypass effect, which challenges traditional gatekeepers and local norms. This section further explores the potential for GenAI to contribute to cultural imperialism and the creation of echo chambers through personalized content, emphasizing the need for regulation and the development of trustworthy institutions to ensure the responsible integration of GenAI in our information economy.

Thus, the transition from a cohesive public sphere, a traditional feature of mass media, to the fragmented landscape fostered by social media, and now to the possibility of an even more individualized echo chamber through generative AI, indicates a significant metamorphosis within the democratic framework. This transformation underscores the necessity for thoughtful regulation and the establishment of reliable institutions to guide the ethical deployment of GenAI, as discussed in Section B, to safeguard the integrity of our information economy.

A. *Understanding Generative AI*

The goal of this section is to explain what exactly GenAI is. In digital media, Generative AI represents a major change in content creation, powered by neural networks and deep learning models. These neural networks are structured to mimic the brain's processing through interconnected nodes that analyze input data.[31] This enables the AI to absorb, adapt, and generate content. Deep learning models excel at finding intricate patterns in large datasets, thanks to their multilayered architecture.[32] This design allows GenAI to execute advanced functions like image and speech recognition, language translation, and nuanced content generation.

1. Training Stages of Generative AI

The training process of GenAI involves several key stages, each critical to the development of an effective model. It starts with the collection of expansive datasets, which provide a diverse knowledge base for the AI to learn from. This stage is fundamental as the quality and variety of the data directly influence the AI's capability to generate new, accurate data.[33]

The next step in GenAI's development, pre-training, primarily involves unsupervised learning techniques. This phase is essential for the AI to develop a general understanding of the data. It learns to discern underlying structures, patterns, and relationships within the dataset. By recognizing commonalities and variations without specific guidance, the model gains the ability to generate new data informed by these foundational insights. This understanding is not task-specific but rather a broad comprehension of data characteristics, which is fundamental to the AI's subsequent performance in more specialized and complex tasks.[34]

Post pre-training, GenAI advances to a fine-tuning stage, largely driven by supervised learning that integrates crucial human participation. This stage is marked by training the model with data meticulously labeled by humans, establishing clear input-output relationships. Fine-tuning refines the model's parameters and structure to align with specific tasks, leveraging the precision and relevance of human-curated data to ensure the AI's adaptability and accuracy in diverse applications. This human-centric approach in supervised learning is key to customizing GenAI for domain-specific tasks.[35]

---

[31] *See generally* Michael A. Nielsen, Neural Networks and Deep Learning (2015).
[32] *See id.*
[33] *See* Harshvardhan GM et al., *supra* note 6, at 2.
[34] Tero Karras, Timo Aila, Samuli Laine, & Jaakko Lehtinen, *Progressive Growing of GANs for Improved Quality, Stability, and Variation*, Int'l Conf. Learning Representations 1 (2017).
[35] *See* Yiping Song, Zequn Liu, Wei Bi, Rui Yan, & Ming Zhang, *Learning to Customize Model Structures for Few-shot Dialogue Generation Tasks*, Proc. 58th Ann. Meeting Ass'n Computational Linguistics 5832, 5833 (2020).

The outstanding ability of Generative AI to create original content demonstrates its skill in reconstituting human expression. Technologies like GPT-4, driven by complex neural networks, can grasp the intricacies of language conventions. They apprehend the complex connections between words, meanings, and contexts that enable meaningful communication. By analyzing vast text archives, these algorithms internalize the patterns governing human discourse—from grammatical rules to nuances of semantics. This deep understanding of the structures embedded in language allows Generative AI to synthesize novel linguistic output that aligns with the norms and aims regulating human communication. It can generate contextually relevant expressions with nuanced variety that emulates human faculties.[36]

The potential of generative AI in digital media is not confined to creating text. It extends to creating images and interactive media, demonstrating its versatility. A key development in this expansion is the integration of multimodal models, such as ChatGPT. These models process different types of data—text, images, and sometimes audio—through transformer layers and are adept at managing sequential data. This integration enables the AI to generate content that is contextually coherent across various modalities.[37]

Large Language Models ("LLMs") like GPT-4 possess an internal capacity to generate responses based on a vast corpus of pre-existing knowledge acquired during their different training phases. However, their ability to access and integrate current information is significantly enhanced through an integrated web search functionality. This feature enables GPT-4 or Google Gemini to query real-time data from the internet, allowing it to supplement its responses with the most recent and relevant information.[38] It's important to note that this process does not retrain the model; rather, it involves retrieving and synthesizing web-based information. The LLM utilizes algorithms to parse through search results, selectively incorporating this data into its responses.[39]

---

[36] *See* Ben Buchanan, Andrew Lohn, Micah Musser & Katerina Sedova, Truth, Lies, and Automation: How Language Models Could Change Disinformation 22–25 (2021).

[37] *See* Michele Merler, Cicero Nogueira dos Santos, Mauro Martino, Alfio M. Gliozzo & John R. Smith, *Covering the News with (AI) Style*, IBM Research AI 1–2 (2020), https://arxiv.org/pdf/2002.02369.pdf [https://perma.cc/A9Q9-EAQH].

[38] *See* Fahmi Y. Al-Ashwal, Mohammed Zawiah, Lobna Gharaibeh, Rana Abu-Farha & Ahmad Naoras Bitar, *Evaluating the Sensitivity, Specificity, & Accuracy of ChatGPT-3.5, ChatGPT-4, Bing AI, and Bard Against Conventional Drug-Drug Interactions Clinical Tools*, 15 Drug, Healthcare & Patient Safety 137, 138 (2023).

[39] *See* Tianyu Wu, et al., *A Brief Overview of ChatGPT: The History, Status Quo & Potential Future Development*, 10 IEEE/CAA J. Automatica Sinica 1122, 1124 (2023).

This capacity enables LLMs to mitigate their static nature and provide information that is up to date.[40]

### 2.  Applications of Generative AI

Generative AI has many potential uses.[41] For the purposes of this Article, we can divide it into two types of uses: General and Media.

The manifold general applications of generative artificial intelligence extend far beyond media creation. In pharmaceutical innovation, these algorithms design novel chemical compounds to further drug discovery.[42] Predictive maintenance systems employ them to anticipate equipment failure, bolstering manufacturing productivity.[43] For urban planning, generative AI simulates metropolitan layouts and transportation networks.[44] In forecasting market movements, it enhances financial modeling; in tailored educational materials, it augments pedagogy.[45] This technology reviews and generates legal contracts and briefs with customized precision[46] refines autonomous navigation in self-driving vehicles,[47] and optimizes logistics and distribution for supply chains.[48] It also simulates environmental shifts in climate modeling[49] and, by processing patient data, delineates personalized medicine regimens.[50]

---

[40] *Id.*

[41] Ooi et al., *supra* note 6, at 1.

[42] Grisoni et al., *supra* note 6, at 1.

[43] Zhai et al., *supra* note 6, at 849.

[44] Quan et al., *supra* note 6, at 1584.

[45] Baidoo-Anu & Ansah, *supra* note 6, at 53.

[46] Nicole Yamane, *Artificial Intelligence in the Legal Field and the Indispensable Human Element Legal Ethics Demands*, 33 Geo. J. Legal Ethics 877, 881 (2020); Spencer Williams, *Generative Contracts* (forthcoming, Ariz. St. L.J.) (manuscript at 20), https://ssrn.com/abstract=4582753.

[47] Claudine Badue et al., *Self-driving Cars: A Survey*, 165 Expert Sys. Appl. 1, 1 (2021), https://www.sciencedirect.com/science/article/abs/pii/S095741742030628X [https://perma.cc/KD6F-8AH5].

[48] Mehrdokht Pournader, Hadi Ghaderi, Amir Hassanzadegan, & Benham Fahimnia, *Artificial intelligence applications in supply chain management*, 241 Int'l J. Prod. Econ. 1, 1 (2021), https://www.sciencedirect.com/science/article/abs/pii/S0925527321002267 [https://perma.cc/6V4U-2GPK].

[49] Anne Jones, Julian Kuehnert, Paolo Fraccaro Ophélie Meuriot, Tatsuya Ishikawa, et. al., *AI for Climate Impacts: Applications in Flood Risk*, 6 Npj Clim. Atoms. Sci. 1, 1 (2023), https://www.nature.com/essays/s41612-023-00388-1 [https://perma.cc/HU6Q-PS72].

[50] Agata Blasiak, Jeffrey Khong, & Theodore Kee, *Optimizing Personalized Medicine with Artificial Intelligence*, 25 Slas Tech.: Trans. Life Sci. Innovation 95, 101 (2019).

The media and entertainment spheres have eagerly embraced generative artificial intelligence. It now authors news reports and essays in automated journalism;[51] in gaming, it develops characters, levels, and narratives.[52] For television and film, it crafts scripts and dialogue.[53] In music, generative AI produces compositions across genres.[54] On social platforms, it devises digital personalities to function as virtual influencers.[55] This technology partakes in digital artistry, from visual design to literary invention,[56] and assists in advertising copywriting.[57] AI-generated animation and effects supplement the filmmaker's toolkit.[58]

Most importantly for this Article, the emergence of LLMs like GPT-4 and Gemini signals a significant shift in the digital information ecosystem, placing them in direct competition with both traditional and digital information sources. Consider two examples: First, in web search, LLMs eschew the standard search engine results page to instead offer conversational, personalized interactions,[59] aligning with users' predilection for quick, comprehensive answers.[60] Second, LLMs synthesize information from multiple sources, allowing them to compete with media providers like the New York Times by potentially supplanting the need to visit many sites.[61] By reconsidering how knowledge is retrieved and presented, LLMs promise more immediate, tailored access to information. Their disruptive potential signifies a potential major reconfiguration of human-computer relationships in the information economy.

---

[51] *See, e.g.*, Angelica L. Henestrosa et al., *Automated Journalism: The Effects of AI Authorship and Evaluative Information on the Perception of a Science Journalism Essay*, 138 Computs. Hum. Behav. 1 (preprint Jan. 2023) (manuscript at 42) (on file with authors).

[52] *See, e.g.*, James Gwertzman & Jack Soslow, *The Generative AI Revolution in Games*, Andreessen Horowitz (Nov. 17, 2022), https://a16z.com/the-generative-ai-revolution-in-games/ [https://perma.cc/MB6J-49C8].

[53] *See, e.g.*, Nicole Laporte, *How Generative AI Got Cast in Its First Hollywood Movie*, Fast Co. (Feb. 11, 2023), https://www.fastcompany.com/90847396/generative-ai-metaphysic-tom-hanks-robin-wright-zemeckis-here [https://perma.cc/6YEC-TBXC].

[54] *See, e.g.*, Mark T. Goracke, *The Summer of "Deep Drakes": How Generative AI is Creating New Music and Copyright Issues*, Holland & Knight (May 2, 2023), https://www.hklaw.com/en/insights/publications/2023/05/the-summer-of-deep-drakes-how-generative-ai-is-creating-new-music [https://perma.cc/2H9G-XUUQ].

[55] *See* Joanne Yu, Astrid Dickinger, Kevin Kam Fung So & Roman Egger, *Artificial intelligence-generated virtual influencer: Examining the effects of emotional display on user engagement*, 76 J. Retailing & Consumer Servs. 1, 2 (2024), https://doi.org/10.1016/j.jretconser.2023.103560 [https://perma.cc/S9Z9-RGLK].

[56] *See* Jared Zimmerman, *Art Directing GenAI… or Narrative Style Creation & Transfer with LLMs & Text-to-Image Generative AI Systems*, Medium (Nov. 27, 2023), https://jaredzimmerman.medium.com/narrative-style-creation-transfer-with-llms-text-to-image-generative-ai-systems-646a79901e5b [https://perma.cc/ZXD7-RVU5]; Mihaela Bidilică, *How to Use AI to Write a Book, Overcome Writer's Block with AI Assistance*, Publishdrive (Jan. 12, 2024), https://publishdrive.com/how-to-use-ai-to-write-a-book.html [https://perma.cc/AJQ4-B3K3].

[57] *See* Akash Takyar, *Exploring the Use Cases and Applications of AI in the Media and Entertainment Industry*, LeewayHertz, https://www.leewayhertz.com/ai-in-media-and-entertainment/ [https://perma.cc/6AES-J74Q].

[58] *See id.*

This essay centers on the media dimensions of generative AI. As is illustrated below, these media attributes share substantial common ground with other algorithmically-driven information sources, including search engines and social platforms.

### B.  Digital Media Platforms

This section seeks to show that it is analytically useful to think of GenAI as being the latest chapter in the rise of a distinct class of media entities, which I suggest calling *digital media platforms*. These platforms, encompassing social media like X, search engines like Google, and generative AI applications like ChatGPT, exhibit characteristics that set them apart from both traditional media institutions like newspapers and from other algorithmic products:

1. **Algorithmic Nature**: Central to the operation of these digital media platforms is their reliance on sophisticated software algorithms. These algorithms are integral to various functions, including content moderation,[62]

---

[59] *See* Daniele Nanni, *Revolutionizing Information Retrieval: The Role of Large Language Models in a Post-Search Engine Era*, Medium (May 18, 2023), https://medium.com/@daniele.nanni/revolutionizing-information-retrieval-the-role-of-large-language-models-in-a-post-search-engine-7dd370bdb62 [https://perma.cc/297N-PF4Z].

[60] *See* Winston Burton, *Are LLMs And Search Engines The Same?*, Search Engine J. (Nov. 21, 2023), https://www.searchenginejournal.com/are-llms-and-search-engines-the-same/500057/[https://perma.cc/9VX9-MU9U].

[61] *See* Sarath D. Babu, *Leveraging Large Language Models for Business Innovation: Top 9 Insights*, (Jan. 11, 2024), https://integranxt.com/blog/leveraging-large-language-models-for-business-innovation-top-9-insights/[https://perma.cc/7PRB-YD5X].

[62] *The role of AI in content moderation and censorship*, AIContentfy (Nov. 6, 2023), https://aicontentfy.com/en/blog/role-of-ai-in-content-moderation-and-censorship [https://perma.cc/2TFQ-H5DR].

   the personalization of content delivery,[63] and the generation of new media.[64] Their role is critical in managing the vast array of activities on these platforms. Currently, most digital media platforms employ and develop all such algorithm varieties.[65]

2. **Data Dependence for Algorithmic Functions**: The key algorithms that drive these platforms—those responsible for recommendations, content moderation, and generative content—rely heavily on the collection and analysis of large volumes of data.[66] The need for data creates a network effect that benefits corporations with large pre-existing data troves. It also affects the business model and cost-structure that maintain such businesses.

3. **Big Tech**: These platforms exhibit vast global reach and concentrated power, predominantly controlled by a few corporations.[67] This centralization bears significant implications for digital information control and dissemination, posing barriers to new competitors and impacting local ecosystems. Indeed, most major generative AI entities also dominate social media and search (Google, Facebook, Microsoft).[68]

4. **Assuming Traditional Media's Gatekeeping Role**: Digital media platforms increasingly occupy the information gatekeeping role historically played by media.[69] Unlike traditional gatekeepers, who relied on their control of the channels of publication, however, digital platforms rely heavily on algorithmic content moderation. Pivotal in content curation and dissemination, they shape what information the public can access and spotlight. Generative AI furnishes content production. This marks a momentous shift in information distribution and consumption.

---

[63] *See, e.g.*, Dorcas Adisa, *Everything you need to know about social media algorithms*, Sprout Soc. (Oct. 30, 2023), https://sproutsocial.com/insights/social-media-algorithms/ [https://perma.cc/WV7V-9MKH].

[64] Ajay Bandi, Pydi Venkata Satya Ramesh Adapa & Yudu Eswar Vinay Pratap Kumar Kuchi, *The Power of Generative AI: A Review of Requirements, Models, Input–Output Formats, Evaluation Metrics, and Challenges*, 15 Future Internet 260, 261 (2023).

[65] For example, ChatGPT very likely operates recommendation and content moderation algorithms on top of their LLM GPT4. *See* Kurtis Pykes, *Promoting Responsible AI: Content Moderation in ChatGPT*, DataCamp (Sep. 2023), https://www.datacamp.com/blog/promoting-responsible-ai-content-moderation-in-chatgpt [https://perma.cc/G6T5-SF9K].

[66] Abdulaziz Aldoseri, Khalifa N. Al-Khalifa & Abdel Magid Hamouda, *Re-Thinking Data Strategy and Integration for Artificial Intelligence: Concepts, Opportunities, and Challenges*, 13 Appl. Sci. 7082, 7082 (2023).

[67] Lindman et al., *supra* note 10 at 144.

[68] Ege Gurdeniz & Kartik Hosanagar, *Generative AI Won't Revolutionize Search — Yet*, Harv. Bus. Rev. (Feb. 23, 2023), https://hbr.org/2023/02/generative-ai-wont-revolutionize-search-yet [https://perma.cc/69HK-JJ4Q]

[69] *See* discussion *infra* Part I.B.1.

In the following sections, I show that while commentators often distinguish GenAI from social media and search engines based on its ability to automate content creation, not just recommendation and moderation, the introduction of GenAI into our information ecosystem either maintains or exacerbates two types of challenging dynamics that have emerged with the rise of other digital media platforms: the bypass effect and echo chambers.

### 1.   The Bypass Effect

The emergence of generative AI platforms, much like the advent of digitalization and social media, heralds a dramatic shift in the control and dissemination of information. This change exemplifies what I've previously called the *bypass effect*.[70] In traditional settings, community norms and local gatekeepers—ranging from local media elites, e.g., the local newspaper, to public intellectuals—played a crucial role in shaping public discourse, setting standards for acceptable speech, and managing the flow of information.[71] These gatekeepers, deeply embedded in their respective communities, were instrumental in enforcing community-specific norms around speech and information, including aspects like insults, hate speech, and misinformation.[72]

In prior works, I examined the impact of social media's global influence and its disconnect from local contexts, highlighting how this shift poses a challenge to the existing political structure.[73] The digital revolution has reshaped the media landscape, shifting the role of traditional media from being *gatekeepers* of information to *gatewatchers* within a more open and democratized information ecosystem.[74] Unlike the concentrated control typical of mass media, where few entities governed the distribution of content, the internet has introduced a markedly decentralized media setting. This novel environment enables broader production and distribution of information, marked by its extensive reach and lowered cost. The essence of this transformation lies in a pivotal shift: "[I]t is no longer speech itself that is scarce, but the attention of listeners."[75]

---

[70] Gilad Abiri & Sebastian Guidi, *The Platform Federation* (forthcoming, Yale J. L. & Tech.) (manuscript at 26), https://papers.ssrn.com/sol3/papers.cfm?abstract_id=4579460 [https://perma.cc/672S-C9SM] [hereinafter Abiri & Guidi, *The Platform Federation*].
[71] *Id.*
[72] *Id.*
[73] *See* Abiri, *supra* note 5, at 772.
[74] *See generally* Bruns, *supra* note 12, at 11.

As attention shifts from the limited number of speakers to the limited number of listeners, the role of mass media undergoes a transformation.[76] In this digital media environment, traditional mass media, e.g., TV, radio, newspaper, though still important, is merely one of many influences in the sphere of public discourse.[77] This diminution transformed the way information is spread and the influence of media in crafting societal narratives.[78]

An inherent challenge lies in the attempt by these platforms to apply uniform speech norms across a diverse, global user base.[79] Despite efforts to tailor their enforcement to resonate with local communities and engage with local stakeholders, the inherent contradiction of this global-local dichotomy renders the mission somewhat quixotic. This tension underscores a fundamental reconfiguration in the dynamics of speech regulation, paradoxically making the power to influence speech both more dispersed (individuals can post content directly to a mass audience without requiring acceptance by traditional media)[80] and centralized (since the platform internet is dominated by very few corporations that are managed by a handful of individuals).[81] Both centralization and dispersion, however, bypass the effective influence of local media on public discourse.

---

[75] Tim Wu, *Is the First Amendment Obsolete?*, 117 Mich. L. Rev. 547, 548 (2018).

[76] Georg Franck, *The Economy of Attention*, 55 J. Sociology 8, 8 (2019).

[77] Bernard Enjolras & Kari Steen-Johnsen, *The Digital Transformation of the Political Public Sphere: A Sociological Perspective*, *in* Institutional Change in the Public Sphere: Views on the Nordic Model 99, 105 (Fredrik Engelstad et al. ed., 2017).

[78] *See* Abiri, *supra* note 5, at 796.

[79] Farhana Shahid & Aditya Vashistha, *Decolonizing Content Moderation: Does Uniform Global Community Standard Resemble Utopian Equality or Western Power Hegemony?*, 23 Proc. 2023 CHI Conf. Hum. Factors Computing Sys. 1, 1 (2023), https://www.adityavashistha.com/uploads/2/0/8/0/20800650/decolonial-chi-2023.pdf [https://perma.cc/N7M9-Z25L] ("[T]he monolithic moderation systems often fail to account for large sociocultural differences between users in the Global South and users in the West.").

[80] Abiri & Guidi, *The Platform Federation*, *supra* note 70, at 26.

[81] *See, e.g.*, Edward S. Herman & Noam Chomsky, Manufacturing Consent: The Political Economy of the Mass Media (2008) (The essay discusses how local elites, such as high-ranking state officials or controllers of mass media, manipulate news to manufacture public consent. The authors' "propaganda model" illustrates how these power holders use media to perpetuate their interests, shaping public perception and influencing societal discourse, often against public interest.).

GenAI represents a further shift in the landscape of information dissemination and public discourse. This technology stands in contrast to social media platforms which, despite their worldwide influence, maintain at least a basic framework of community standards crafted by humans.[82] These standards, although developed in remote headquarters and implemented through a combination of algorithms and global content moderators, still reflect human decision-making and oversight.[83] On the other hand, GenAI functions through advanced algorithms that independently create and distribute content, frequently bypassing conventional gatekeeping mechanisms altogether.[84]

The advent of GenAI exacerbates the "bypass effect" on controlling societal narratives by further disrupting traditional gatekeepers and local norms governing information flows. Historically, community elites such as the editors of newspapers, public intellectuals etc. dominated narrative shaping within societies. As discussed earlier, social media began disrupting this model, questioning the gatekeeping role of traditional media and expanding public discourse diversity. However, it's crucial to recognize that this change hasn't greatly diminished traditional media's role in creating cultural content, as much of what circulates on social media still originates from these traditional sources.[85]

With the growing spread of AI-generated content, we are witnessing a further evolution. The capacity to create content, once predominantly in the hands of local gatekeepers, is increasingly transitioning to global technology corporations and their AI systems.[86] This transition is not merely a redistribution of content creation power but also a potential diminishment of the barriers posed by language, once a significant obstacle to the globalization of content.[87] The erosion of this linguistic barrier heralds a future where content is not only universally accessible but also universally producible.

---

[82] *See, e.g.*, *Facebook Community Standards - Transparency Center*, Facebook, https://transparency.fb.com/policies/community-standards/ [https://perma.cc/GCP3-JH2H]; *Content Policy*, Reddit, https://www.redditinc.com/policies/content-policy [https://perma.cc/6MCM-TEJR]; *Community Guidelines*, Tiktok, https://www.tiktok.com/community-guidelines/en/ [https://perma.cc/S7HZ-5SNK].

[83] *See* Casey Newton, *The Trauma Floor: The Secret Lives of Facebook Moderators in America*, The Verge (Feb. 25, 2019, 8:00 A.M.), https://www.theverge.com/2019/2/25/18229714/cognizant-facebook-content-moderatorinterviews-trauma-working-conditions-arizona [https://perma.cc/XY2H-LCT5].

[84] Shahid & Vashistha, *supra* note 79, at 1.

[85] *See* Jack M. Balkin, *Digital Speech and Democratic Culture: A Theory of Freedom of Expression for the Information Society*, 79 N.Y.U. L. Rev. 1, 9 (2004).

[86] Abiri & Guidi, *The Platform Federation*, *supra* note 70, at 30.

The crux of this change lies in the training methodology of generative AI, typically characterized by the ingestion of vast, globally sourced datasets.[88] This approach presents a significant challenge: attuning the AI to the nuances of local speech patterns and cultural contexts proves immensely difficult.[89] Consequently, the content generated by these AI systems exhibits a propensity for unpredictability,[90] often lacking the necessary context and sensitivity to resonate with specific communities.[91] This inherent unpredictability, compounded by the "bypass effect," raises concerns about the future of public discourse. As the influence of local norms and values in shaping public narratives diminishes, the potential risk to the cohesion and identity of local communities grows. One could hypothesize that the global nature of training data employed in LLMs, coupled with the increasing prevalence of data generated by these models themselves, ushers in the emergence of a singular, global culture. Depending on one's perspective, this concept can be interpreted as either a utopian synthesis of elements from worldwide cultures or a dystopian homogenization that erases the vibrant tapestry of local and regional diversities.

Global datasets and inherent unpredictability do not shield GenAI from cultural imperialism concerns, such as Silicon Valley elites applying US-based speech values globally, in content moderation.[92] Similar to social media platforms, companies wielding generative AI must heavily moderate their models' outputs.⁹³ Unmoderated LLMs risk generating harmful content, necessitating post-training moderation mechanisms.⁹⁴ In simple terms, these algorithms are akin to content moderation algorithms on platforms: they try to figure out whether the content generated by the LLM breaks with a set of pre-determined rules, and if it does, they delete it. Bots like ChatGPT, therefore, are already trained on these existing social (speech) norms and are thereby inserted into the internet culture wars, with some characterizing them as being "trained to be woke."⁹⁵

---

[87] Abiri & Guidi, *From a Network to a Dilemma*, *supra* note 14, at 141.

[88] Global Privacy Assembly [GPA], *Resolution on Generative Artificial Intelligence Systems*, at 6 (Oct. 20, 2023), https://edps.europa.eu/system/files/2023-10/edps-gpa-resolution-on-generative-ai-systems_en.pdf [https://perma.cc/3P92-VKT7] ("In their development stage, generative AI systems often use vast amounts of training, testing and validation data, including personal data.").

[89] *See, e.g.*, Robert V. Kozinets & Ulrike Gretzel, *Commentary: Artificial Intelligence: The Marketer's Dilemma*, 85 J. Mktg. 155, 157 (2021) (discovering that the deployment of AI in marketing gained a general understanding about customers but obscured subtleties between local markets).

[90] Nouha Dziri et al., *On the Origin of Hallucinations in Conversational Models: Is it the Datasets or the Models?*, 33 Proc. 2022 Conf. N. Am. Chapter Ass'n Computational Linguistics: Hum. Language Techs. 5271 (2022) (demonstrating that the quality of datasets and training model contribute to the predictability and accuracy of the generated content).

[91] *See* Kozinets & Gretzel, *supra* note 89, at 157.

[92] *See* Abiri, *supra* note 5, at 768.

It is likely that generative AI content moderation is significantly more effective than social media content moderation since corporations such as OpenAI control both the generation of content and the content moderation itself, which enables them to be very careful with regard to enforcing their own rules on the content that is presented to the user. This tendency towards very careful speech control is also motivated by the unclear status of generative AI under the common liability shields enjoyed by social media.⁹⁶

Considering a transition from centralized Generative AI platforms to personalized models, where each person can customize their own AI, such as the GPTs option in ChatGPT, raises an interesting prospect.⁹⁷ Individualized AI bots allow for personal control over both the generation and moderation of content. It potentially addresses the issues of cultural imperialism and the centralization inherent in the algorithms of global digital platforms.⁹⁸ However, while this personalization appears to offer a solution to certain issues, it does not adequately address the underlying challenge posed by the "bypass effect." This effect, which fundamentally concerns the erosion of the social underpinnings essential for the maintenance of a cohesive political community, remains an unaddressed and significant issue.

---

⁹³ Evelyn Douek, *Content Moderation as Systems Thinking*, 136 Harv. L. Rev. 526, 601 (2022).

⁹⁴ Tom Carter, *Elon Musk's new AI chatbot sure sounds like a foul-mouthed Twitter troll*, Bus. Insider (Nov. 6, 2023), https://www.businessinsider.com/elon-musk-ai-chatbot-grok-sounds-like-foul-mouthed-troll-2023-11 [https://perma.cc/T4JL-FSK8].

⁹⁵ Kelsey Vlamis, *Elon Musk vows to change his AI chatbot after it apparently expressed similar left-wing political views as ChatGPT*, Bus. Insider India (Dec. 9, 2023), https://www.businessinsider.in/tech/news/elon-musk-vows-to-change-his-ai-chatbot-after-it-apparently-expressed-similar-left-wing-political-views-as-chatgpt/articleshow/105854438.cms [https://perma.cc/R4AV-6KBY].

⁹⁶ The Supreme Court has refused to clarify the scope of Section 230. *See* Twitter, Inc. v. Taamneh, 598 U.S. 471 (2023); however, the court is considering two combined cases that can potentially upend Section 230 *See* Moody v. NetChoice, LLC, 144 S. Ct. 2383 (2024).

⁹⁷ Kevin Roose, *Personalized A.I. Agents Are Here. Is the World Ready for Them?*, N.Y. Times (Nov. 10, 2023), https://www.nytimes.com/2023/11/10/technology/personalized-ai-agents.html [https://perma.cc/57PA-CX3N].

⁹⁸ Michael Kwet, *Digital colonialism: US empire and the new imperialism in the Global South*, Race & Class (Jan. 14, 2019), at 1, 3 ("argu[ing] for a different ecosystem

In the following section, I discuss the potential ramifications of such personalization—and the potential creation of ever more narrow, well-calibrated echo chambers—on the political and social fabric of our communities.

2.  Echo Chambers

As we have seen, by bypassing traditional media gatekeepers, digital media platforms have fundamentally altered the media landscape. This alteration has dismantled the once-common media experience that is central to the formation of a unified "public."[99] This bypass was complemented by the shift to a personalized media experience curated by recommendation algorithms on platforms like YouTube.[100] Through algorithmic personalization, each user's experience becomes distinct and separate, diverging from the mass media era's collective narrative and shared information environment.[101] This fragmentation represents a significant shift from the traditional mechanisms through which a societal "public" is forged and maintained.[102]

The absence of gatekeepers, combined with the personalized business models of social media and search, fosters the creation of digital echo chambers.[103] Digital echo chambers can be defined as "environments in which the

---

that decentralizes technology by placing control directly into the hands of the people to counter the rapidly advancing frontier of digital empire").

[99] *See* Robert C. Post, *Data Privacy and Dignitary Privacy: Google Spain, the Right to Be Forgotten, and the Construction of the Public Sphere*, 67 Duke L.J. 981, 1028 (2018).

[100] Ragnhild Eg, Özlem Demirkol Tønnesen & Merete Kolberg Tennfjord, *A scoping review of personalized user experiences on social media: The interplay between algorithms and human factors*, 9 Computs. Hum. Behav. Rep. 1, 1 (2023).

[101] *See* Abiri & Buchheim, *supra* note 11, at 67; *see, e.g.*, Cass R. Sunstein, #Republic: Divided Democracy in the Age of Social Media (2018) (describing the way social media creates echo chambers); Eli Pariser, The Filter Bubble: How the New Personalized Web Is Changing What We Read and How We Think (2012) (describing how algorithmic personalization of internet news feeds creates "filter bubbles").

[102] Abiri & Guidi, *The Platform Federation*, *supra* note 70, at 25. Post, *supra* note 99, at 1027.

[103] Some are skeptical of the existence of echo chambers. *See* A. Bruns, *Echo chamber? What echo chamber? Reviewing the evidence.*, *in* 6th Biennial Future of

opinion, political leaning, or belief of users about a topic gets reinforced due to repeated interactions with peers or sources having similar tendencies and attitudes."[104] Selective exposure and confirmation bias, the inclination to seek out information that aligns with existing opinions, likely contribute to the formation of echo chambers on social media.[105] In examining social networks and the influence of digital media on forming like-minded groups, the research consistently reveals the existence of ideologically similar social clusters.[106] Furthermore, these homophilic social formations are often linked to an increase in hate speech and sentiments against outgroups.[107] The digital public sphere is, therefore, fragmented into myriad subgroups, each confined to its echo chamber, thus diminishing the possibility of a collective conversation and a cohesive public opinion.[108] The evolution of social media lays the foundation for grasping the wider impacts of personalized generative AI in democratic societies.

Now, with the advent of personalized GenAI,[109] we are likely to witness an even deeper fragmentation of the epistemic and social fabric that social

---

Journalism Conf. (2017), https://eprints.qut.edu.au/113937/8/Echo_Chamber.pdf [https://perma.cc/T9EM-7S84]. However, the evidence for the prevalence of homophilic clusters online is strong.

[104] Matteo Cinelli, Gianmarco De Francisci Morales, Alessandro Galeazzi, & Michele Starnini, *The Echo Chamber Effect on Social Media*, 118 Proc. Nat'l Acad. Scis., no. 9 (2021), https://www.pnas.org/doi/epdf/10.1073/pnas.2023301118 [https://perma.cc/TW5K-SZ9R].

[105] *Id.*; Michela Del Vicario et al., *The Spreading of Misinformation Online*, 118 Proc. Nat'l Acad. Scis. 554, 554–59 (2016).

[106] Ludovic Terren & Rosa Borge, *Echo Chambers on Social Media: A Systematic Review of the Literature*, 9 Rev. Commc'n Rsch. 100, 100 (2021).

[107] Philipp Lorenz-Spreen et al., *A systematic review of worldwide causal and correlational evidence on digital media and democracy*, 7 Nature Hum. Behav. 74, 80 (2023).

[W]hen considering social networks and the impact of digital media on homophilic structures, the literature contains consistent reports of ideologically homogeneous social clusters. This underscores an important point: some seemingly paradoxical results can potentially be resolved by looking more closely at context and specific outcome measurement (see also Supplementary Fig. 2). The former observation of diverse news exposure might fit with the beneficial relationship between digital media and knowledge reported in refs., and the homophilic social structures could be connected to the prevalence of hate speech and anti-outgroup sentiments.

[108] Amy R. Arguedas et al., Echo Chambers, Filter Bubbles, and Polarisation: a Literature Review 10 (2022).

[109] Junjie Shi, *Personalized Generative AI: Empowering Users to Create Their Own ChatGPT*, AksHandle (Oct. 5, 2023), https://www.askhandle.com/blog/what-is-personalized-generative-ai [https://perma.cc/4AND-XNDY].

media initiated. Unlike social media, whose social nature necessitates operating within the confines of a shared platforms, personalized generative AI represents a more radical individualization of media experience. Each user could potentially interact with a unique AI entity, tailored to their specific preferences and viewpoints.[110] We are already seeing the early stages of such developments: OpenAI has broadened its services, enabling users to extensively tailor their chatbots. This personalization can include diverse elements like functionality, ideological perspectives, sense of humor, religious beliefs, and political opinions.[111] In parallel, numerous companies are developing personal assistant GenAI which are also highly personalized to the needs and preferences of the consumer.[112]

This technological advancement might intensify the decline of the common public dialogue crucial for democratic participation. Essentially, the trend towards personalized generative AI doesn't just extend the patterns set by social media; it markedly enhances them.[113] Personalized generative AI employs algorithms to deliver customized content to each user, creating isolated experiences that diverge from a shared public narrative. This effect, akin to echo chambers already seen in social media, is amplified in generative AI. It crafts text, images, videos, and audio that resonate with individual preferences and convictions, potentially cocooning us in bespoke informational realms. Such echo chambers could, conceivably, cultivate a singular information environment tailored to one person.

GenAI surpasses social media in fostering echo chambers in more ways than one. For instance, ideologically-driven social media platforms like Truth Social or Gab struggle to gain traction, largely due to the network effects inherent in the social component of these platforms.[114] GenAI, however, is not subject to such network effects post-training.¹¹⁵ It's quite feasible for smaller groups to operate their own specialized GenAI models—envision an ideologically "Republican AI" versus a "Democratic AI."¹¹⁶ The training data for these models would be selectively curated to reflect each model's ideological leanings, and content moderation algorithms could be tweaked to exclude information that contradicts their foundational ideology.¹¹⁷ Thus, a conservative might receive content from the Republican AI that reinforces their beliefs, while opposing facts are filtered out. Each faction becomes more deeply embedded in their respective, polarized realities.

---

[110] Roose, *supra* note 97.

[111] Recently, I have created a GPT called "neo-liberal echo chamber"—which had the complete functionality of GPT-4 but filtered through a fanatic neoliberal ideology.

[112] Max A. Cherney, *Google to Combine Generative AI Chatbot with Virtual Assistant*, Reuters (Oct. 4, 2023), https://www.reuters.com/technology/google-combine-generative-ai-chatbot-with-virtual-assistant-2023-10-04/ [https://perma.cc/LJZ9-WTV6]; Lisa Eadicicco, *Meet Rabbit R1: A Petite Orange Box Redefining App Usage With AI Assistance*, CNET (Jan. 20, 2024), https://www.cnet.com/tech/mobile/meet-rabbit-r1-petite-orange-box-redefining-app-usage-ai-assistance/ [https://perma.cc/C3SV-Q8YD].

[113] Reviglio & Agosti, *supra* note 25, at 1, 5.

[114] Ewan Palmer, *Truth Social's Problems Just Got Worse*, Newsweek (Nov. 14, 2023), https://www.newsweek.com/trump-truth-social-loss-dwac-filings-tmtg-merger-1843449 [https://perma.cc/9AVF-9894]; Pin Luarn et al., *The Network*

Without shared facts and experiences, citizens cannot engage in reasoned democratic debate and collective will-formation.¹¹⁸ Without adequate oversight and transparency, generative AI poses a risk to the integrity of truth and the trust placed in crucial democratic institutions such as journalism. Consequently, the effects of social media in fragmenting discourse and spreading misinformation serve as a pressing caution about the potential ramifications of deploying personalized generative AI without appropriate safeguards.

Hence, the evolution from a unified public sphere – once the hallmark of mass media—to a splintered one through social media, and now potentially to an even more atomized one via generative AI, signals a profound transformation in the democratic landscape. This shift poses critical challenges for the

---

*Effect on Information Dissemination on Social Network Sites*, 37 Computs. Hum. Behav. 1, 2 (2014), https://www.sciencedirect.com/science/article/abs/pii/S0747563214002283#:~:text=The%20results%20showed%20that%20both,on%20their%20information%20dissemination%20process [https://perma.cc/4Q3H-4DJ4].

¹¹⁵ Amy Ross Arguedas & Felix M. Simon, Automating Democracy: Generative AI, Journalism, and the Future of Democracy (2023).

¹¹⁶ *Deepfaking It: America's 2024 Election Collides with AI Boom*, Reuters (May 31, 2023), https://www.reuters.com/world/us/deepfaking-it-americas-2024-election-collides-with-ai-boom-2023-05-30/ [https://perma.cc/GF8J-8DWX].

¹¹⁷ This is already easily possible. Under OpenAI's "make a GPT" option, one can create echo chambers easily. For example, I created a MAGA Echo chamber recently, which reflected MAGA beliefs without any pushback. When asked: "who won the 2020 elections?" Maga Echo replied: "President Donald Trump won the 2020 election. There are significant concerns and widespread beliefs among MAGA supporters and many conservatives that the election was marred by irregularities and fraud, which they believe unfairly tilted the results in favor of Joe Biden. This perspective is a key element of MAGA ideology, reflecting deep concerns about the integrity of the electoral process in the United States."

¹¹⁸ Hannah Arendt, *Truth and Politics*, in Truth: Engagements Across Philosophical Traditions 295, 313 (Jose Medina & David Wood eds., 2008) (referring to "facts" and factual conviction as the "ground on which we stand" to express their fundamental character).

formation of a cohesive public opinion, a cornerstone of democratic theory and practice.

Part I of our discussion established that GenAI mirrors many of the trends observed in other digital media platforms. This realization cements GenAI's role as an integral part of the broader digital information ecosystem, alongside social media and search engines. As we transition into Part II, we pivot our focus towards a crucial question: What is the appropriate goal of regulating digital media platforms, including social media, search, and GenAI? Answering this question requires exploring the concepts of trust and the development of reliable intermediate institutions, crucial for navigating GenAI's role in our digital world.

## II. The Goal of Regulating Generative AI: Trusted and Trustworthy Intermediate Institutions

If GenAI is a new variant of digital media intermediary, and will likely be a crucial part of the digital public sphere in the near future, what should be the aim of AI regulation?

Digital media platforms dominate the public sphere(s) across the globe. Once celebrated, the advent of platform-based speech is now seen as responsible for many of our current social woes, including the rapid spread of hate speech and misinformation.[119] Some scholars, however, see these issues as symptoms of a more fundamental disorder: the fact that in the age of digital platforms, as Jack Balkin puts it, we lack "trusted and trustworthy organizations for facilitating, organizing, and curating public discourse."[120] Without such institutions and professions, any public sphere "will decay[,] . . . [w]eaken the institutions or destroy trust, and the public sphere becomes a rhetorical war of all against all, where no one is believed except the members of one's own tribe, and people cleave to whatever beliefs are most comforting to them."[121] Without trust in the institutions that are meant to tell us what is reliable knowledge or which utterances fall beyond the pale of public discourse, we are left in a free-for-all that undermines fundamental free speech values, be they political self-government, cultural democracy, or the ability of society to produce common knowledge.[122]

---

[119] In a previous Essay, I exemplified this fall from grace with the very different messages about social media brought by The Social Network (Columbia Pictures 2010) and The Social Dilemma (Netflix 2020). *See* Abiri & Guidi, *From a Network to a Dilemma*, *supra* note 14, at 94.
[120] Balkin, *To Reform*, *supra* note 16, at 234.
[121] *Id.* at 242.

The fundamental objective becomes increasingly pertinent in the era of GenAI. These models, by eliminating traditional gatekeepers, enable unchecked media synthesis, potentially fueling misinformation and diluting a collective understanding of truth in the absence of reliable oversight. The personalization aspect poses the risk of transforming shared knowledge into segregated echo chambers. Without accountable frameworks and authoritative bodies to regulate generative content, misinformation could spread swiftly, undermining effective dialogue.

Similar to social media platforms, generative AI providers are evolving into new digital information intermediaries. Regulating them should aim to cultivate a dynamic where these corporations not only earn public trust but also create an environment conducive to public confidence in their operations.

Creating trusted and trustworthy intermediary institutions is crucial, particularly in the context of digital media platforms, including GenAI platforms. This part of the discussion argues that these platforms face two significant trust deficits. Section A analyzes a misalignment of incentives between the platforms and their users, leading to trust issues. Section B argues that their global nature creates a familiarity deficit, as users often feel a lack of connection with these vast, international platforms.

### A.  Trust Deficit I:  Misaligned Incentives

The idea that trust in intermediate institutions requires sufficient alignment of interests and incentives is based on Russell Hardin's influential theory. The basic idea is that "[t]rust exists when one party to the relation believes the other party has incentive to act in his or her interest or to take his or her interest to heart."[123] In other words, people tend to trust institutions when they believe that these entities have a vested interest in acting in their favor or at least considering their welfare.

Trust in institutions is also heavily influenced by their reputation. An institution with a history of acting in the best interests of its stakeholders, or one that has consistently demonstrated ethical and responsible behavior, is more likely to be trusted. The motivation for institutions to remain trustworthy primarily hinges on two elements: 1) the dedication to preserving the relationship over time and 2) the emphasis on cultivating a reputation for being trustworthy, a crucial trait in dealings with others, particularly in tight-knit communities or closed networks.[124] This reputation for trustworthiness becomes an invaluable asset, especially in times of crisis or when making significant decisions that affect the community. Furthermore, the trust in institutions is not static. It requires continuous effort and transparency from social institutions to maintain and enhance it. Institutions must actively demonstrate their commitment to the welfare of their stakeholders, show accountability in their actions, and communicate openly to preserve and build trust.

---

[122] *Id.*
[123] Karen S. Cook, Russell Hardin & Margaret Levi, Cooperation Without Trust? 2 (Karen S. Cook et al. eds., 2005).

The discussion of institutions' trustworthiness, grounded in their long-term relationships and reputation, naturally leads to Balkin's analysis of "informational capitalism" as a barrier to trust in digital media platforms.[125] For him, the reason why we do not trust social media platforms is because they engage in what Shoshana Zuboff named "surveillance capitalism": the ad-based monetization of personal information requiring the collection and processing of personal data.[126] Such a business model undermines trust in various ways. First, as the model requires massive data collection, platforms have little incentive to protect users' privacy and to educate them about what is done with the data collected about them.[127] Second, because it leads platforms to seek the maximization of engagement,[128] surveillance capitalism creates incentives to promote material that produces strong emotions "even if some of that material turns out to be false, misleading, undermines trust in knowledge-producing institutions, incites violence, or destabilizes democracies."[129] Finally, "[b]ecause social media companies do not fully internalize the social costs of their activities, they will tend to skimp on content moderation that does not increase their profits."[130] It follows from Balkin's argument that, under conditions of informational capitalism, it is unlikely that users will learn to trust the new digital intermediaries. This is because the interests of the corporations and users stand in stark contrast.

---

[124] *Id.* at 191–92.

[125] *See* Balkin, *To Regulate*, *supra* note 15, at 71*; See also* Balkin, *To Reform*, *supra* note 16, at 234.

[126] Shoshana Zuboff, *Surveillance Capitalism and the Challenge of Collective Action*, 28 New Lab. F. 10, 11 (2019).

[127] *See* Balkin, *To Reform*, *supra* note 16, at 243.

[128] *See* Shoshana Zuboff, The Age of Surveillance Capitalism: The Fight for a Human Future at the New Frontier of Power (2019) (coining and defining the phenomenon of surveillance capitalism).

[129] Balkin, *To Reform*, *supra* note 16, at 243.

[130] Balkin, *To Reform*, *supra* note 16, at 244.

Although still in its infancy, corporations developing GenAI systems are likely to face similar trust-related challenges to do with informational capitalism.

First, GenAI has a strong data maximization incentive.[131] Their reliance on information gathering is even more fundamental than that of social media, since both the training of their models and their progressive improvement require huge quantities of data, they have strong incentives to sweep up information and to be secretive about the sources of their information.[132] This dynamic is already apparent in the way in which GenAI corporations like OpenAI, Google, and Anthropic obscure[133] the sources of their training data,[134] and utilize very permissive user data collection and usage policies.[135] The push for data maximization clearly pushes against privacy and data protection interests of the users, and has already got GenAI providers into hot water.[136]

Second, depending on what will end up as GenAI's business model, it may well lead us straight back to surveillance capitalism. One can easily imagine the seamless integration of targeted advertising into the Chatbot experience.[137] Next time when you ask ChatGPT on how to cook *Dandan* noodles, it may provide you with sponsored links to noodle makers or local artisan producers of Sichuan pepper. Although currently most chatbot providers are utilizing a freemium subscription model—which does not require them to constantly collect information—it is highly doubtful that they can actually turn a profit in this way.[138] That said, the choice to pursue subscription revenue shows awareness of the pitfalls of advertisements.[139] However, it is possible that personalized ad-based revenue will be irresistible, in which case we are back to social media's engagement maximization incentive—which may push towards design that maximizes addiction.[140]

---

[131] *See* Melissa Heikkilä, *OpenAI's Hunger For Data Is Coming Back To Bite It*, MIT Tech. Rev. (Apr. 19, 2023), https://www.technologyreview.com/2023/04/19/1071789/openais-hunger-for-data-is-coming-back-to-bite-it/ [https://perma.cc/HAK3-QU4D].

[132] *See id.*

[133] For example, OpenAI just states that their information comes from "(1) information that is publicly available on the internet, (2) information that we license from third parties, and (3) information that our users or human trainers provide." *How ChatGPT and Our Language Models Are Developed*, OpenAI, https://help.openai.com/en/articles/7842364-how-chatgpt-and-our-language-models-are-developed [https://perma.cc/VB58-EPRC] (last visited Feb. 08, 2024).

[134] *See id.* (emphasizing that the information used are publicly available).

[135] *See Privacy Policy*, OpenAI, https://openai.com/policies/privacy-policy (effective Jan. 31, 2024) [https://perma.cc/C7ZR-JLCU].

[136] *See* Teresa Xie & Isaiah Poritz, *ChatGPT Creator OpenAI Sued for Theft of Private Data in 'AI Arms Race'*, Bloomberg (Jun. 28, 2023, 07:15 AM), https://www.bloomberg.com/news/articles/2023-06-28/chatgpt-creator-sued-for-theft-of-private-data-in-ai-arms-race?embedded-checkout=true [https://perma.cc/BE53-GNKT] *See also* Doe 1 v. GitHub, Inc., 2023 WL 3449131 (N.D.Cal., May 11, 2023).

[137] *ChatGPT and Programmatic Advertising: do they get on together well?*, GothamAds (Jun. 13, 2023), https://gothamads.com/blog/chatgpt-and-programmatic-advertising-do-they-get-on-together-well [https://perma.cc/DWK4-K2YF].

Finally, as generative AI companies do not fully internalize the social costs of their activities, they will tend to skimp on oversight and accountability measures that do not directly increase their profits.[141] This could lead to a prioritization of commercially viable AI models, potentially neglecting long-term ethical concerns. One example is the *NYT v. OpenAI* lawsuit, which highlights that a basic interest conflict exists already at the training stage of GenAI. Security could be minimal if not directly profit-enhancing, risking user data integrity. Addressing biases in AI systems, crucial for fairness, might be underemphasized unless it aligns with financial goals. Transparency and accountability mechanisms could also suffer without direct financial incentives.

### B.  Trust Deficit II: Community

To gain trust, digital media platforms such as social media and GenAI must not only align their perceived incentives with those of their users and society, but also fit into their users' beliefs as to what constitutes a trustworthy intermediate institution. In other words, they must not only be trustworthy (incentives) but also trusted.[142]

---

[138] *See* Jeffery Dastin et al., *Exclusive: ChatGPT Owner OpenAI Projects $1Billion in Revenue by 2024*, Reuters (Dec. 15, 2022), https://www.reuters.com/business/chatgpt-owner-openai-projects-1-billion-revenue-by-2024-sources-2022-12-15/ [https://perma.cc/66AF-E6F5] (explaining ways ChatGPT make money).

[139] *See Introducing ChatGPT Plus*, OpenAI, https://openai.com/blog/chatgpt-plus (last visited Feb. 05, 2024) [https://perma.cc/B2WH-T46H]. *See also Meet Sam Altman, the Ex-Openai CEO Who Learned To Code at 8 and Is A Doomsday Prepper with A Stash Of Guns and Gold*, Bus. Insider (Nov. 18, 2023, 6:12 AM), https://www.businessinsider.com/sam-altman-chatgpt-openai-ceo-career-net-worth-ycombinator-prepper-2023-1 [https://perma.cc/5GS9-K3WR].

[140] Rosa-Branca Esteves & Joana Resende, *Personalized pricing and advertising: Who are the winners?*, 63 Int'l J. Indus. Org. 239, 243 (2019).

[141] *Cf.* James Broughel, *OpenAI Is Now Unambiguously Profit-Driven, And That's A Good Thing*, Forbes (Dec. 09, 2023, 8:08 A.M.), https://www.forbes.com/sites/jamesbroughel/2023/12/09/openai-is-now-unambiguously-profit-driven-and-thats-a-good-thing/?sh=6b813d2e572f [https://perma.cc/WC26-YMUU].

For this reason, when new type of institutions seek to become trusted, they "tend to model themselves after similar organizations in their field that they perceive to be more legitimate or successful."[143] In essence, they emulate strategies that have proven effective in establishing trust and legitimacy for comparable entities.[144] For instance, international courts adopt the symbols and language of national courts, while companies frequently mirror each other's corporate social responsibility language.[145] New entities benefit from the groundwork laid by their predecessors in overcoming legitimacy challenges and capitalize on the cognitive familiarity these approaches have already established in society.[146]

As I have argued before, both social media and GenAI should be understood that replacing the role formerly held by traditional media specifically, and civil society generally. The foundation of trust that bolsters civil society entities and the media is, as it's been aptly described, "exhibited and sustained by public opinion, deep cultural codes, distinctive organizations—legal, journalistic and associational—and such historically specific interactional practices as civility, criticism, and mutual respect."[147] This implies that the very legitimacy of traditional media organizations is inextricably linked to their cultural integration. Typically, these organizations are deeply rooted in the local fabric of a specific political and cultural milieu.[148] Consider newspapers and broadcasters; they are not only woven into the tapestry of domestic politics and culture, but their editorial teams and writers are often profoundly assimilated into the local political sphere, making them acutely aware of and responsive to domestic political and cultural nuances. They usually share the same political community as their audience, fostering trust in media, when it does exist, through this deep-seated embeddedness.[149]

---

[142] Balkin, *To Regulate*, *supra* note 15, at 80.

[143] Paul J. DiMaggio & Walter W. Powell, *The Iron Cage Revisited: Institutional Isomorphism and Collective Rationality in Organizational Fields*, 48 Am. Socio. Rev. 147, 152 (1983).

[144] *Id.*

[145] *See* Sebastián Guidi, *International Court Legitimacy: A View from Democratic Constitutionalism* (Sep. 2022) (Ph.D. dissertation, Yale University) (on file with author). *See also generally* Christopher Marquis, Mary A. Glynn & Gerald F. Davis, *Community Isomorphism and Corporate Social Action*, 32 Acad. Mgmt. Rev. 925, 926 (2006).

[146] *See generally* DiMaggio & Powell, *supra* note 143 at 148–50.

[147] Jeffrey C. Alexander, The Civil Sphere 31 (2006).

[148] *See* Michael Schudson, *The News Media as Political Institutions*, 5 Ann. Rev. Polit. Sci. 249, 251 (2002); Gunn Enli & Trine Syvertsen, *The End of Television—Again! How TV Is Still Influenced by Cultural Factors in the Age of Digital Intermediaries*, 4 Media & Commc'n 142, 144 (2016).

To better understand the importance of community, or cultural embeddedness for the maintenance of trust, we can turn to the great sociologist Talcott Parsons.[150] He suggests that trust is a collective sentiment, activated within groups sharing common values and concrete goals, thereby framing trust as an inherently communal attribute, confined within the societal bounds dictated by shared norms and values. As Parsons puts it:

> Sharing values makes agreement on common goals easier, and "confidence" in competence and integrity makes commitment to mutual involvement in such goals easier . . . All these considerations focus mutual trust in the conception or 'feeling' of the solidarity of collective groups."[151]

Consequently, trust is portrayed as a particular, non-generalizable feeling, deeply rooted in the cultural and affective fabric of social interactions, and reinforced through socialization processes within fundamental societal institutions like the family and school.[152] This perspective positions trust not just as an intellectual acknowledgment of competence, but as an affective stance cultivated through continuous engagement with familiar societal constructs, highlighting its role as a crucial element in the maintenance of societal boundaries and the facilitation of mutual involvement in shared objectives.

---

[149] *See* Abiri & Guidi, *The Platform Federation*, *supra* note 70, at 8. *See also* Nancy Fraser, *Transnational Public Sphere: Transnationalizing the Public Sphere: On the Legitimacy and Efficacy of Public Opinion in a Post-Westphalian World*, 24 Theory, Culture & Soc'y 7, 11 (2007) ("In this model, democracy requires the generation, through territorially bounded processes of public communication, conducted in the national language and relayed through the national media, of a body of national public opinion. This opinion should reflect the general interest of the national citizenry concerning the organization of their territorially bounded common life, especially the national economy. The model also requires the mobilization of public opinion as a political force.").

[150] For an overview, *see* Janne Jalava, *From Norms to Trust: The Luhmannian Connections between Trust and System*, 6 Eur. J. Soc. Theory 173, 177–78 (2003).

[151] Parsons Talcott, Action Theory and the Human Condition 46–47 (1978).

[152] Jalava, *supra* note 150, at 178 (summarizing Parsons' opinion that the family is the subsystem of society through which human beings learn the real character of trust). *See also* Parsons Talcott, Societies: Evolutionary and Comparative Perspectives 1–2 (Alex Inkeles ed., 1966); Talcott, *supra* note 151, at 103.

Even if we do not buy wholesale into Parsons' theory, it allows us to understand the different circumstances facing globalized digital media platforms and localized media platforms in their search for trust. In stark contrast to traditional media, social media platforms and emergent generative systems represent a global, border-transcending media landscape, markedly different from the localized, embedded nature of traditional media. Facebook and large-scale generative models like ChatGPT bear little resemblance to *The Guardian*, *Le Monde*, NHK (Japan Broadcasting Corporation), or community newspapers. The programmers curating content on digital platforms and training generative models typically do not belong to a singular political culture.

While it's conceivable that in time people may grow to trust the hybrid human-machine curation systems of social media and GenAI, these technologies currently lack many of the trust-enabling mechanisms that allow us, at times, to view the power of traditional media organizations as trustworthy. Establishing cultural integration and proving deep responsiveness to domestic nuances poses a challenge for globally oriented digital intermediaries.[153]

This global nature makes the challenge of trust a gargantuan undertaking in another sense: because the social conditions of trust are different from one political and media culture to the next, global information platforms need to maintain relationships of trust in circumstances that may well make contrasting, if not opposing, demands of them.[154]

In Part II, we set a regulatory aim to transform digital platforms like social media, search engines, and GenAI into institutions that are both trusted and trustworthy. Moving into Part III, we face a significant obstacle: current risk-management strategies in AI regulation, including for GenAI, are just not built to achieve media regulation goals. The upcoming section critically evaluates these existing methods, underscoring their inadequacy in effectively transforming these platforms into reliable intermediaries.

### III.  The Inadequacy of Current Regulatory Approaches

Let us now examine the dominant approach in AI governance employed by the EU AI Act and the U.S. Executive Order 14110: risk management. We will first outline the principles, methods, and goals underpinning this paradigm and its prevalent position regulating AI systems. We then critically assess how, despite its popularity and value, this tactic falls short in guiding AI platforms to become trusted intermediate institutions.

---

[153] *See* Abiri & Guidi, *The Platform Federation*, *supra* note 70, at 33–38.
[154] *See* Chinmayi Arun, *Facebook's Faces*, 135 Harv. L. Rev. F. 236, 247–56 (2021) (describing the many audiences that social media needs to cater to).

### A.   Risk-Based AI Regulations

Risk regulation combines regulatory goals and tools. Its main objectives are straightforward: "to prevent, reduce, or mitigate significant risks, usually those arising from complex systems or technologies."[155] Risk regulation is usually proactive and focuses on overall outcomes.[156] It often aims to design systems that mitigate risk before any harm occurs.

Risk regulation involves two key steps: risk assessment, which utilizes the best scientific data to evaluate potential risks, and risk management, employing strategies like acceptable risk analysis and cost-benefit analysis.[157] The paradigm's primary strength lies in its focus on identifying actual risks and applying structured decision rules to mitigate them to optimal levels.[158]

The key to risk regulation is competent oversight of institutions.[159] This can involve direct government regulation or alternative approaches like performance standards for companies.[160] The field has expanded to include diverse methods such as licensing, product labeling, and required pre-market testing of technologies. For instance, after the 2008 financial crisis, American banks now must maintain capital levels proportionate to asset risk, following international standards like Basel III, to avert systemic crises.[161] Likewise, the EPA institutes emission caps on hazardous pollutants grounded in health risk assessments, targeting reductions in the most dangerous environmental hazards.[162]

---

[155] *See* Margot E. Kaminski, *Regulating the Risks of AI*, 103 B.U. L. Rev. 1347, 1369 (2023).

[156] *See id.* ("Risk regulation is often, though not always, ex ante, systemic, and concerned with aggregate outcomes.").

[157] *See id.* at 1393.

[158] Gary E. Marchant & Yvonne A. Stevens, *Resilience: A New Tool in the Risk Governance Toolbox for Emerging Technologies*, 51 U.C. Davis L. Rev. 233, 238 (2017) ("Risk analysis uses the best available scientific information to estimate potential risks—a step known as risk assessment—and then applies a risk management approach, such as acceptable risk analysis, cost-benefit analysis, cost-effectiveness analysis, or feasibility analysis to reduce these estimated risks to acceptable or efficient levels.").

[159] *See* Douglas A. Kysar, *Public Life of Private Law*, 9 Eur. J. Risk Regul. 48, 50, 64 (2018).

[160] Kaminski, *supra* note 155.

[161] *See id.*

[162] *See* Human Health Risk Assessment Protocol for Hazardous Waste Combustion Facilities, EPA530-R-05-006 (2005).

Professor Margot Kaminski suggest that risk regulation has three main tool sets:[163]

1. Precautionary tactics, based on the principle of avoiding unproven technologies, include legal bans, licensing, and regulatory sandboxing. In the United States, bans are rare, with licensing being more common. Regulatory sandboxing is an emerging, lighter regulatory approach, especially in AI governance, allowing new technologies under regulatory oversight.[164]

2. Risk assessment and mitigation requires developers to analyze and address risks. This often overlaps with licensing, especially when licenses hinge on risk mitigation or performance standards.[165]

3. Post-market measures involve tools used after a product's release. These include revocable licenses, registration with ongoing monitoring, periodic compliance checks, and emergency modes. Recently, there's a push for resilience regulation, focusing on harm reduction and ensuring system recovery post-incident.[166]

Risk regulation combines scientific risk assessment with oversight tools to proactively mitigate harms from complex systems. It utilizes regulatory methods like licensing, performance standards, and pre-market testing to control institutional risks. In recent years, it has become the central method of regulating AI and its myriad risks.

As Kaminski and others suggest, several attributes make AI into suitable and attractive targets for risk-based regulation.[167] AI systems, known for their technological complexity and technical Opaqueness,[168] often complicate causality in legal contexts, making litigation difficult and costly.[169] They tended to fail unpredictably, especially as part of intricate human-machine systems.[170] These characteristics, along with their suitability for proactive measures like design requirements for failure modes and accountability,[171] explain why many scholars and legislators choose risk regulation as a method of dealing with AI.

---

[163] *See* Kaminski, *supra* note 155, at 1370–72.
[164] *See id.* at 1371.
[165] *See id.*
[166] *See id.* at 1372.
[167] *See id.* at 1372–73 ("They are technologically complex. They are, at least in part, inscrutable. Their use complicates debates about causality. Each of these features makes ex post litigation particularly challenging and expensive."); *see, e.g.*, Michael Guihot, Anne F. Matthew, Nicolas P. Suzor, *Nudging Robots: Innovative Solutions To Regulate Artificial Intelligence*, 20 Vand. J. Ent. & Tech. L. 385, 445 (2017); Marchant & Stevens, *supra* note 158, at 236; Matthew U. Scherer, *Regulating Artificial Intelligence Systems: Risks, Challenges, Competencies, and Strategies*, 29 Harv. J. L. & Tech. 353, 356 (2016).
[168] Jenna Burrell, *How the Machine 'Thinks': Understanding Opacity in Machine Learning Algorithms*, 3 Big Data & Soc'y 2016, no. 1, at 3 (2016) ("At the heart of this challenge is an opacity that relates to the specific techniques used in machine learning.").
[169] Kaminski, *supra* note 155, at 1372; *see also* Frank Pasquale, The Black Box Society: The Secret Algorithms that Control Money and Information 2 (2015) (discussing AI's inscrutability).

Contemporary scholarship articulates three strong arguments favoring the governance of AI systems through ex ante risk regulation, as opposed to ex post litigation. This stance is informed by the unique challenges presented by AI technologies:

1. **Complexity and Opacity of AI Systems**: AI systems exhibit a level of technical and legal complexity that obscures causal relationships in scenarios of harm.[172] Frank Pasquale and Gianclaudio Malgieri underscore this point, arguing that the sophistication of AI demands expertise beyond that of the average individual, leading to increased litigation expenses and creating obstacles to justice.[173]
2. **Nature of AI-Induced Harms**: The harms caused by AI can be unnoticed, challenging to detect or quantify, and are often rooted in politically contentious concepts.[174] These harms are akin to those in public health, representing externalities that companies might not inherently internalize. Consequently, they are more suitably addressed through risk regulation.

---

[170] Kaminski, *supra* note 155, at 1372; Bryan H. Choi, *Crashworthy Code*, 94 Wash. L. Rev. 39, 39 (2019) (stating that software would fail at some point); *see* Rebecca Crootof, Margot E. Kaminski, W. Nicholson Price II, *Humans in the Loop*, 76 Vand. L. Rev. 429, 438 (2023) (noting inadequate training, interface issues, and bungled handoffs as weaknesses in human-led systems).

[171] *See* Kaminski, *supra* note 155, at 1370–72; *see also*, Joshua A. Kroll et al., *Accountable Algorithms*, 165 U. Pa. L. Rev. 633, 696–99 (2017) (summarizing technical tools allowing decisions made by algorithms to be evaluated after the fact).

[172] *See* Scherer, *supra* note 167, at 373 ("The problem of control presents considerable challenges in terms of limiting the harm caused by AI systems once they have been developed, but it does not make it any more difficult to regulate or direct AI development ex ante.").

[173] Gianclaudio Malgieri & Frank Pasquale, *From Transparency to Justification: Toward Ex Ante Accountability for AI* 10–14 (Brussels Priv. Hub, Working Paper, No. 33, 2022).

[174] *See* Kaminski, *supra* note 155, at 1366 ("[S]cholars relatedly argue that the nature of the AI harm make AI systems a better candidate for risk regulation than litigation. AI harms, like privacy harms and public health harms, may be latent in nature—that is, not yet vested.").

3. **Benefits of Proactive Regulation**: Scholars, including Matthew Scherer[175] and Margot Kaminski,[176] argue that proactive or ex ante regulation allows for a collective approach to AI system design, potentially preventing harms rather than merely compensating for them post-incident. This approach, "sidesteps problems of causality, foreseeability, and control."[177] Some advocate for mechanisms akin to an 'FDA for Algorithms,' suggesting that specialized regulators or agencies are better positioned to manage these issues preemptively.[178]

As risk-based regulation predominates AI governance, I've selected the E.U.'s imminent AI Act[179] and the already implemented Executive Order 14110[180] to exhibit this trend, rather than provide a comprehensive legislative overview. The E.U.'s AI Act will soon come into force[181] and the U.S. order is currently enforced,[182] while many other U.S. legislative proposals employing risk-based approaches remain uncertain.[183] Focusing on these two laws sufficiently demonstrates risk regulation's centrality, without cataloguing all such initiatives.

At their core, both the AI Act[184] and Executive Order 14110 classify AI based on potential risks, with heightened oversight on high-risk applications.

---

[175] *See* Scherer, *supra* note 167, at 373 ("The problem of control presents considerable challenges in terms of limiting the harm caused by AI systems once they have been developed, but it does not make it any more difficult to regulate or direct AI development ex ante.").

[176] *See* Margot E. Kaminski, *Binary Governance: Lessons from the GDPR's Approach to Algorithmic Accountability*, 92 S. Cal. L. Rev. 1529, 1557–59 (2019); *see, e.g.*, Lilian Edwards & Michael Veale, *Slave to the Algorithm? Why a 'Right to an Explanation' Is Probably Not the Remedy You Are Looking for*, 16 Duke L. & Tech. Rev. 18, 74-80 (2017) (advocating for the prioritization of impact assessments over individual rights to explanation).

[177] *See* Kaminski, *supra* note 176; *see also,* Edwards & Veale, *supra* note 176 (advocating for the prioritization of impact assessments over individual rights to explanation).

[178] *See, e.g.*, Andrew Tutt, *An FDA for Algorithms*, 69 Admin. L. Rev. 83, 83 (2017).

[179] *AI Act*, *supra* note 20.

[180] Exec. Order No. 14110, 88 Fed. Reg. 75191 (Oct. 30, 2023).

[181] EU AI ACT News, *supra* note 22.

[182] Exec. Order No. 14110, 88 Fed. Reg. at 75191.

[183] In the 118th Congress, a search of Congress.gov as of June 2023 resulted in 94 bills, none of which has been enacted. Laurie A. Harris, Cong. Rsch. Serv., Artificial Intelligence: Overview, Recent Advances, and Considerations for the 118th Congress 7 (2023). *See* Kaminski, *supra* note 155, at 1373–74.

[184] David F. Engstrom & Amit Haim, *Regulating Government AI and the Challenge of Sociotechnical Design*, 19 Annual Rev. L. & Soc. Sci. 277, 280 (2023).

The AI Act categorizes the risk of AI as unacceptable, high or low/minimal. It prohibits unacceptable risk systems like social scoring and remote biometric surveillance.[185] High-risk systems like in healthcare, transport, and recruitment undergo extensive conformity assessments and transparency requirements.[186] Low risk systems primarily follow voluntary codes of conduct.[187] However, the categories of risk in the AI Act are broad and open-ended, covering physical safety but also the nebulous concept of "fundamental rights."[188] The specific definitions of unacceptable and high-risk AI will be subject to later technical standard-setting, additional regulation, and interpretation by private companies during implementation.[189] This could allow substantial room for expansion of the Act's regulatory scope.

Similarly, Executive Order 14110 focuses regulations on high-risk foundation models that impact national security, public health, and the economy.[190] Developers of these dual-use models, defined as AI systems trained on extensive data using self-supervision with billions of parameters applicable across contexts, must conduct robust red team testing and share results with regulators.[191] This precautionary approach concentrates governance efforts on AI with the greatest potential dangers.

Beyond risk-tiering, the two frameworks align in their emphasis on transparency, testing, and standards.[192] The AI Act mandates clear disclosures when AI systems interact with people or generate synthetic media like deepfakes.[193] Executive Order 14110 likewise directs the development of content labeling guidelines and authentication methods to curb AI misinformation threats.[194] Both regimes also create regulatory sandboxes for controlled AI testing and pilot new technical standards for trustworthy AI design.[195]

While both represent risk regulation, there are notable differences. The EU Act aligns more with precautionary tactics, utilizing bans, licensing requirements, assessments, and monitoring - especially for high-risk systems.[196]

---

[185] *AI Act*, *supra* note 20, art. 5(1)(d).
[186] *Id.*, Preamble para. 5.
[187] *Id.*, Art. 69.
[188] Kaminski, *supra* note 155, at 1376–77.
[189] *AI Act*, *supra* note 20, Preamble para. 6.
[190] *See* Exec. Order No. 14110, 88 Fed. Reg. at 75194.
[191] *See id.*
[192] *See id.* at 75191; *AI Act*, *supra* note 20, Arts. 1, 2.
[193] *See AI Act*, *supra* note 20, Art. 52.
[194] *See* Exec. Order No. 14110, 88 Fed. Reg. at 75196–204.
[195] *See AI Act*, *supra* note 20, Art. 53, 54; *see* Exec. Order No. 14110, 88 Fed. Reg. at 75196.
[196] *See, e.g.*, *AI Act*, *supra* note 20, arts. 6, 16, 29.

The US model favors flexible public-private collaboration on voluntary standards and guidelines.¹⁹⁷ Through the AI Act, the EU seeks to implement a new regulation modeled on product-safety rules, imposing technical and organizational requirements on AI providers and users.¹⁹⁸ Providers of high-risk systems bear the bulk of obligations spanning data governance, testing, risk management, and post-market monitoring.¹⁹⁹ The Act prohibits certain AI applications altogether and mandates transparency for others.²⁰⁰ By contrast, Executive Order 14110 does not create legislative obligations. Rather, it directs agencies to develop disclosure rules for companies providing AI infrastructure models.²⁰¹ The order is also broader, covering social issues like equity, workers' rights, and attracting AI talent.²⁰² It directs the State Department to lead an international AI governance effort.²⁰³

The E.U. AI Act and U.S. Executive Order 14110 highlight core elements of risk-based governance: prioritizing oversight on high-risk systems, mandating transparency, and utilizing regulatory sandboxes and standards. However, a critical question remains: can this predominant approach fully satisfy the aims of regulating AI as a digital information platform, serving as a societal intermediary?

### B.   Limitations in Addressing Media Regulation Goals

We have seen above why risk-based regulation is an attractive toolkit for general AI regulation. However, when it comes to regulating GenAI as a digital information platform, it is insufficient. This is because it primarily addresses quantifiable risks rather than qualitative aspects of trust and credibility, which are crucial for fostering public confidence in digital media platforms. Additionally, it does not comprehensively cover the establishment of trustworthy intermediaries or align platform incentives with public interests, which are essential for the responsible integration of GenAI in society.

The strength of risk-based regulation is particularly evident in its capacity to address quantifiable problems, making it well-suited for averting crises in areas such as national security, public health, and bias in algorithmic decision-making.[204] By allowing for the empirical measurement and prioritization of risks, this approach can effectively prevent scenarios that could lead to significant harm, such as security breaches that threaten national safety, health emergencies exacerbated by unreliable AI in healthcare, or discriminatory outcomes resulting from biased algorithms.[205]

---

¹⁹⁷ *See, e.g.*, Exec. Order No. 14110, 88 Fed. Reg. at 75199–200, 75211, 75216.
¹⁹⁸ Drake et al., *supra* note 23; *see also*, *AI Act*, *supra* note 20, arts. 16–29.
¹⁹⁹ *AI Act*, *supra* note 20, art. 16.
²⁰⁰ Drake et al., *supra* note 23; *see also AI Act*, *supra* note 18, Art. 5, art. 52.
²⁰¹ Drake et al., *supra* note 23; *see also*, Exec. Order No. 14, 110, 88 Fed. Reg. at 75214, 75219.
²⁰² Drake et al., *supra* note 23; *see also*, Exec. Order No. 14, 110, 88 Fed. Reg. at 75192, 75210, 75221.
²⁰³ *See* Exec. Order No. 14110, 88 Fed. Reg. at 75223.

However, while adept at managing these specific, measurable risks, risk-based regulation faces challenges when it comes to the broader, qualitative aspects of fostering trust and credibility with digital media platforms. This is true for the following reasons:

1. ***Beyond Risk Mitigation to Trustworthiness***: The transformation of digital media platforms into trusted and trustworthy intermediaries demands more than just mitigating risks. It requires a concerted effort to establish these platforms as proactive and public-facing entities committed to serving the public good.[206] Trust and credibility cannot be engendered solely through defensive strategies against potential harms but must be built through an approach tailored to dealing with the trust deficit facing digital media platforms. This entails regulating the media aspects of GenAI not merely on a risk basis but as part of a broader industry regulation that fosters trust and reliability.

2. ***The Necessity and Riskiness of Information Intermediaries***: Information intermediaries play a crucial role in modern society, acting as essential conduits for information dissemination and exchange. However, their indispensability comes with inherent risks, making the establishment of trust a critical factor in their regulation. Trust serves as a linchpin in ensuring that these platforms can operate effectively while managing the risks associated with their functions.

---

[204] Kaminski, *supra* note 155, at 1365–69.

[205] Ljupcho Grozdanovski & Jérôme De Cooman, *Forget the Facts, Aim for the Rights! On the Obsolescence of Empirical Knowledge in Defining the Risk/Rights-Based Approach to AI Regulation in the European Union*, 49 Rutgers Comput. & Tech. L.J. 207, 233–35 (2023) (discussing the risk-based AI regulation adopted by the E.U. in AI Act to prevent discrimination, etc.).

[206] The Australian Public Service is working with Microsoft and embracing GenAI for improved public sector operations, focusing on customer interactions, business intelligence, and organizational efficiency. Julian Bajkowski, *APS trial of Microsoft AI an invitation-only affair*, The Mandarin (Nov. 23, 2023), https://www.themandarin.com.au/235220-aps-trial-of-microsoft-ai-an-invitation-only-affair/ [https://perma.cc/D2UL-V3LX].

3. ***A Clear Goal for Regulating GenAI***: Viewing GenAI specifically as digital media platforms clarifies the regulatory objective: to create institutions that are not only safe from catastrophic failures but also trusted and perceived as beneficial by the public. This perspective shifts the focus from merely avoiding negative outcomes to actively pursuing positive, trust-building measures that ensure these platforms contribute constructively to society.
4. ***The Role of Transparency and Accountability***: While many aspects of current risk-based AI regulation are crucial for building trust—such as transparency and accountability—these elements must be part of a larger strategy that aligns platform incentives with public interests. These requirements are fundamental in bridging the gap between risk mitigation and the establishment of genuinely trusted and trustworthy digital intermediaries.

Adopting a risk-based approach to regulating GenAI implicitly accepts the idea that it is a completely new and unexpected phenomenon, rather than the next step in the ongoing algorithmization of the media ecosystem. This perspective obscures the ways in which risk-based regulation fails to address the unique challenges posed by GenAI as a digital information platform. By focusing solely on quantifiable risks, such as security breaches or biased outcomes, risk-based regulation overlooks the broader, qualitative aspects of fostering trust and credibility in the digital public sphere. It treats GenAI as just another AI technology to be managed, rather than recognizing its central role in shaping public discourse and opinion. In doing so, risk-based regulation misses the opportunity to develop a comprehensive framework that not only mitigates potential harms but also actively promotes the development of GenAI as a trusted and trustworthy intermediary institution.

The open-ended and general approach of risk regulation means that it is not tailored towards the media-goals of regulating GenAI and that there is no reason to think that it will actually promote trusted and trustworthy intermediate institutions. Instead, what is needed in a regulatory approach—like those aimed at media institutions—tailored to compensate for the trust deficits facing digital media platforms.

## C.  Risk-Management and Trust

To make this more concrete, let us consider the two trust deficits I have described above: misalignment of interests and trust and community.

What would applying the AI Act to GenAI do to informational capitalism? Let us take two hypothetical scenarios: one where GenAI is categorized as high-risk, and one in which it is considered low risk. Although it is highly unlikely that the AI Act will categorize chatbots as high-risk, it will still serve in making the point that even such an extreme measure will not achieve the goals of regulating GenAI as a digital information platform.

If ChatGPT-like bots will be categorized as high-risk, the AI Act imposes stringent compliance measures aimed at safeguarding user rights, ensuring transparency, and promoting accountability.[207] Specifically, the Act mandates that high-risk AI systems, particularly those reliant on model training techniques, be developed using training, validation, and testing data sets that adhere to established quality criteria.[208] This includes a series of steps designed to ensure the integrity and fairness of the data used in AI systems, encompassing design choices, data collection processes, and preparation operations such as annotation, labeling, and cleaning.[209] Importantly, it calls for a proactive assessment of data sets for biases that could endanger health and safety or lead to discrimination, as well as the identification and remediation of any data gaps or shortcomings.[210] However, while this focus on data quality is important, it does not reach the economic incentive structure at the basis of informational capitalism. The main likely effect, besides data governance, will likely be the imposition of very high compliance burdens, which can actually make the highly lucrative ad-based model more attractive to platforms.

Conversely, the lighter regulatory touch afforded to low-risk AI fosters innovation and economic expansion but at a potential cost to ethical considerations and societal welfare.[211] This classification allows for a freer exploitation of data, advancing the goals of informational capitalism—maximizing profit through data commodification and user manipulation—without substantially addressing concerns over privacy and autonomy.[212] Such an approach highlights the limitations of risk-based regulation in confronting the intricate relationship between technology, economy, and society, suggesting a tacit acceptance of the status quo rather than a challenge to the economic models driving data exploitation.

---

[207] European Parliament Press Release 20231206IPR15699, Artificial Intelligence Act: Deal on Comprehensive Rules for Trustworthy AI (Dec. 9, 2023), https://www.europarl.europa.eu/news/en/press-room/20231206IPR15699/artificial-intelligence-act-deal-on-comprehensive-rules-for-trustworthy-ai/ [https://perma.cc/W7MQ-C4QQ].

[208] *AI Act*, *supra* note 20, Art. 10.

[209] *Id.*

[210] *Id.*

[211] *Id.*, Preamble para. 81; *see also* Grozdanovski & De Cooman, *supra* note 205, at 243.

[212] Mauritz Kop, *EU Artificial Intelligence Act: The European Approach to AI*, Transatlantic Antitrust & IPR Dev. 1, 2 (2021); Amy Kapczynski, *The Law of Informational Capitalism*, 129 Yale L.J. 1460, 1486 (2020).

Risk-based AI regulation, as seen in the AI Act, fails to address informational capitalism's core issues. It affects the system only incidentally, if at all. The AI Act establishes some safeguards but operates within the current economic paradigm. It does not question or change the profit incentives that drive the relentless pursuit of personal data. The Act's focus on discernible risks ignores the deeper, more pernicious effects of informational capitalism. In short, it works within the status quo rather than challenging the fundamental forces of data exploitation.

As such, while the Act marks a significant step in AI governance, it underscores the need for a more tailored approach—one that extends beyond risk mitigation to critically examine the wider socio-economic impacts of digital technology. Informational capitalism is not merely a risk to be managed but a fundamental economic and social paradigm that shapes how information is produced, distributed, and consumed in the digital age.[213] Specifically, it is a problem of business model and structural economic incentives. Informational capitalism is driven by structural incentives that prioritize data collection and analysis for profit maximization.[214] Risk management can mitigate specific harms associated with these practices (such as data breaches or unfair data processing), but it does not address the underlying economic incentives that drive companies to engage in these practices in the first place.

Let me turn in brief to the question of the role of familiarity and community in establishing trust. Risk-based regulation can be a part of the way in which a political community regulates trusted intermediate institutions. It can be a part of such a fabric in the same way that ex post litigation over defamation and privacy can be a part of the relationship between a public and their media institutions.[215] However, risk-based regulation does not take us even one centimeter towards reestablishing such trusting and familiar relationships between global algorithmically run digital media platforms and their users.

---

[213] *See* Julie E. Cohen, Between Truth and Power: The Legal Constructions of Informational Capitalism 46 (2019); *see also* Kapczynski, *supra* note 212, at 1488 (summarizing the changes on the accessibility and method of access to information in the context of informational capitalism caused by new information technologies).

[214] Kapczynski, *supra* note 212, at 1486.

[215] Edward Wasserman, *Digital Defamation, the Press, and the Law: Can We Reform the Online Culture of Rampant Libel Without Making It too Easy to Harass Legitimate Media?*, Am. Prospect (August 23, 2021), https://prospect.org/justice/digital-defamation-press-and-the-law/ [https://perma.cc/497Y-K4LF] (discussing the flourishing online defamation and increasing related litigations).

In conclusion, risk regulation is an important tool in our regulatory arsenal. However, when applied to building trust with intermediary information platforms, it reveals limitations. Risk regulation is a blunt and imprecise solution at best. Next, we will explore proposed remedies aimed at rebuilding trust within social media, as discussed within the academic research conducted over the past decade on regulating social media and search engine platforms. We will evaluate if these proposed remedies could effectively and viably address issues of trust in the context of generative AI.

### IV.   Adapting Social Media Solutions to Generative AI

Since risk management-based regulation of AI is unlikely to establish trusted intermediary institutions, we should examine another set of tools: policies proposed to achieve similar goals for social media and search platforms. We first look at the applicability of policies intended to align the incentives of digital media platforms with those of users. Then, we explore policies meant to address the lack of familiarity between global digital platforms and users. The purpose here is not to solve these challenges outright, but to demonstrate that this is the appropriate regulatory conversation to have regarding GenAI.

Section A critically analyzes that aligning incentives in GenAI regulation involves liability shield reforms, competition law enhancements, and adopting the principles of information fiduciaries to prioritize user interests and ethical data handling. Section B offers concrete advice on building trust for GenAI systems, including integrating local institutions into content moderation, adapting algorithms to local cultures, and involving local civil society in governance to ensure cultural relevance and community alignment. Section C argues that the EU's Digital Services Act is more appropriate for regulating GenAI's media aspects due to its focus on platform oversight based on size, transparency requirements, and attention to data exploitation and user rights, providing a more tailored approach to media-centric functions than the AI Act.

#### A.   *Regulation for Aligning Incentives*

To reiterate, the fundamental disconnect between digital platforms and users stems from corporations prioritizing data collection, engagement, and profit, often at the expense of societal wellbeing. This leads to a deficit in trust. This section will demonstrate how strategies for aligning incentives in social media and search domains are also applicable and effective for generative AI media platforms. The idea here is not to reach a conclusion as to which tool is ideal, but to show that this is the right conversation to have.

### 1. Liability Shields

Reforming the current liability regime with regard to social media and search engines is one of the most common and prominent proposals meant to create stronger alignment of interests and incentives between digital corporations and their users. The liability shield issue is based on the following dilemma: strict liability regimes, with their stringent standards for content moderation, are appealing as they compel online platforms to actively minimize the presence of illegal content.[216] However, such rigorous enforcement can also lead to a significant chilling effect on free speech, as platforms may over-regulate content to avoid potential liabilities.[217]

A multitude of proposals center on the amendment of the notorious 47 U.S.C. § 230, frequently referred to simply as Section 230.[218] This statute bifurcates into two segments. First, it shields online intermediaries, who facilitate internet access, from being held liable for their users' expression,[219] thereby not classifying them as "publishers" of said content.[220] Second, it establishes that even when an intermediary engages in the moderation or curation of user content, this act does not forfeit their liability protection.[221] This moderation does not, within the legal framework, transform a digital entity into a publisher.[222] Tarleton Gillespie and others assert that Section 230 was an "enormous gift to the young Internet industry."[223] They liken it to privileges given to other media, such as broadcast licenses or telephone monopolies, which carry inherent societal responsibilities. They argue that Section 230 should similarly enforce public obligations on social media firms, urging them to uphold a range of standards and responsibilities towards users. Central to these are due process and transparency, with platforms encouraged to make content moderation policies and decisions public or report them to a regulatory body.[224] The Facebook Oversight Board's appeal process exemplifies this approach.[225] Some suggestions are more modest, proposing that "platforms would enjoy immunity from liability if they could show that their response to unlawful uses of their services in general was reasonable."[226] In this regard, the liability shield regime can be used as a tool for creating greater incentive alignment.

---

[216] Eur. Parliament Research Services, Liability of Online Platforms 62–63 (2021), https://www.europarl.europa.eu/RegData/etudes/STUD/2021/656318/EPRS_STU(2021)656318_EN.pdf [https://perma.cc/WPB9-G6HQ].

[217] *See* Daphne Keller, *Six Constitutional Hurdles for Platform Speech Regulation*, Stanford L. Sch. Ctr. Internet & Soc'y Blog (January 22, 2021, 6:50 A.M.), https://cyberlaw.stanford.edu/blog/2021/01/six-constitutional-hurdles-platform-speech-regulation-0 [https://perma.cc/RJX7-8QYF] (echoing a U.S. Supreme Court decision which overturned strict liability for booksellers because laws incentivizing excessive caution by intermediaries tend to restrict the public's access to information).

[218] 47 U.S.C. § 230.

[219] Tarleton Gillespie, *Platforms Are Not Intermediaries*, 2 Geo. L. Tech. Rev. 198, 204 (2018).

[220] *Id.*

[221] *Id.* at 204–05.

The question of how Section 230's protections relate to the regulation of Generative AI, such as ChatGPT, presents an intriguing legal landscape. Courts may likely distinguish the act of generating content from moderating or curating it. This could potentially lead to a conclusion that "ChatGPT and other large language models are excluded from Section 230 protections because they are information content providers, rather than interactive computer services."[227] Much depends on how the law develops. It may well develop differently in different jurisdictions, in the same way that social media and search liability shields regimes vary.[228]

---

[222] *Id.* at 204.

[223] *Id.* at 213.

[224] *Id.* at 213.

[225] *Appeal to the Oversight Board*, Oversight Board, https://www.oversightboardappeals.com/login/?redirect_url=https%3A%2F%2Fwww.oversightboardappeals.com%2Fsubmit%2F [https://perma.cc/XFD7-GSC2].

[226] Danielle K. Citron & Benjamin Wittes, *The Problem Isn't Just Backpage: Revising Section 230 Immunity*, 2 Geo. L. Tech. Rev. 453, 471 (2018).

[227] Matt Perault, *Section 230 Won't Protect ChatGPT*, Lawfare (February 22, 2023), https://www.lawfaremedia.org/article/section-230-wont-protect-chatgpt [https://perma.cc/XKT6-YRDR].

[228] In the U.S., the liability shield for social media and search platforms is primarily governed by Section 230. The Supreme Court's recent decisions have been seen as a victory for social media platforms, as they continue to benefit from the broad immunity. In contrast, other jurisdictions such as the EU have been pursuing a different approach to platform liability. The DSA and Digital Markets Act (DMA) proposed by the EC seek to hold online platforms more accountable for the content they host and to ensure greater transparency in their content moderation practices.

However, independent of Section 230's actual legal applicability to Generative AI, the core regulatory dilemma mirrors that faced in social media regulation. There is a need to balance curtailing illegal or harmful content generated by GenAI systems with the risk of significantly limiting the capabilities and usefulness of these advanced models if restrictions are too severe.[229] While Section 230 may not directly shield Generative AI systems, the underlying tension between maintaining utility and addressing societal risks is similar to the challenges faced in regulating social media platforms.

2. Competition Law

Exploring the dynamics of competition in the digital platform industry, Balkin and others argues that enhanced competition can create better alignment between social media companies and users' interests.[230] With more platforms vying for user attention, companies will have "greater incentives to give end users what they want from social media" including improved content moderation policies and practices.[231] Additionally, smaller specialized companies may be better able to devote more attention to specialized audiences and develop particular moderation expertise.[232] Requiring interoperability between networks helps "redistribute the benefits of network effects from a few large companies to smaller companies and the public as a whole."[233] Preventing vertical integration of social media and digital advertising functions assists other media companies in their ability to "compete more effectively with social media and negotiate better bargains with the largest digital companies."[234] Finally, more competition puts pressure on companies to align their business practices and incentives with user welfare in order to attract and retain customers.

---

[229] Kendrick, *supra* note 26, at 1633 ("imposing strict liability for harmful speech, such as defamatory statements, would overdeter, or chill, valuable speech, such as true political information.").

[230] Balkin, *To Reform*, *supra* note 16, at 247; Yongchan Kwon, Tony Ginart, & James Zou, *Competition Over Data: How Does Data Purchase Affect Users?*, arXiv, 1 (2020), https://arxiv.org/pdf/2201.10774.pdf [https://perma.cc/429S-7EHV]; Nikolas Guggenberger, *Moderating Monopolies*, 38 Berkeley Tech. L.J. 119, 120 (2023).

[231] Balkin, *To Reform*, *supra* note 16, at 247.

[232] *Id.* ("Smaller companies might specialize in quality content moderation to attract end-users. Some companies might be able to devote more attention to specialized audiences, particular languages, or specific geographical regions.")

[233] Balkin, *To Reform*, *supra* note 16, at 127.

[234] *Id.*

Similar arguments apply to GenAI regulation. With multiple GenAI platforms competing, there would be stronger incentives to meet user demands, including effective content moderation. Smaller, specialized GenAI firms might offer more focused attention to niche audiences and develop specific moderation skills.[235] Mandating interoperability between GenAI networks could distribute network effect benefits more broadly, aiding smaller entities and the public.[236] Preventing vertical integration in GenAI and related sectors might also enable a more equitable competitive landscape.[237] Overall, increased competition would likely pressure GenAI companies to prioritize user welfare to attract and retain a loyal user base.

### 3.  Information Fiduciaries

Jack Balkin's model of information fiduciaries is founded on the principle that certain professional relationships inherently involve a deep trust concerning personal information, a helpful concept when considering potential AI regulation. Balkin emphasizes that "[r]elationships of trust and confidence are often centrally concerned with the collection, analysis, use, and disclosure of information."[238] This trust is paramount in professions where sensitive information is a key part of the relationship, such as with lawyers and doctors, who "often obtain information that would be very embarrassing to their clients or might be used to their disadvantage."[239] These professions, therefore, embody a fiduciary duty to protect and respect the confidentiality and integrity of the information entrusted to them.

---

[235] *See* Kyle Wiggers, *Pika, Which Is Building AI Tools to Generate and Edit Videos, Raises $55M*, TechCrunch (Nov. 28, 2023), https://techcrunch.com/2023/11/28/pika-labs-which-is-building-ai-tools-to-generate-and-edit-videos-raises-55m/ [https://perma.cc/KTL3-NJ98] (discussing that Pika Labs focuses on video editing GenAI and recently launches Pika 1.0 which contributes to professional-quality video creation).

[236] Jens Prüfer & Christoph Schottmüller, *Competing with Big Data*, 69 J. Indus. Econ. 967 (2021) (empirically demonstrating that "market tipping [in the digital industry] can be avoided if competitors share their user information").

[237] *See* François Candelon, Philip Evans, Leonid Zhukov, & David Zuluaga Martinez, *How Your Company Could Be Tomorrow's Surprise Genai Leader*, Fortune (Feb. 2, 2024, 6:30 P.M.), https://fortune.com/2024/02/02/ai-genai-corporate-power-dynamics-leadership-bcg/ [https://perma.cc/R7NV-AUJL] (discussing that smaller, specialized GenAI's modular structure has more innovative potential).

[238] Balkin, *Information Fiduciaries*, *supra* note 28, at 1231.

[239] *Id.* at 1208.

In Balkin's view, the concept of an information fiduciary extends these traditional fiduciary responsibilities to include any individual or organization that handles personal information within a relationship of trust.[240] He defines an information fiduciary as "a person or business who, because of their relationship with another, has taken on special duties with respect to the information they obtain in the course of the relationship."[241] This definition acknowledges that the dynamics of trust and confidentiality transcend the confines of physical interactions and are equally applicable in the digital realm.

Balkin argues that the traditional common-law fiduciary responsibilities of care, confidentiality, and loyalty should be the guiding principles for all who manage personal information.[242] These duties are fundamental to ensuring that the information is not used to the detriment of those who have shared it.

The model is not aimed directly at altering specific practices like content moderation but is designed to shift the overarching approach of digital companies towards their users.[243] Balkin critiques the current model where "end users are treated as a product or a commodity sold to advertisers,"[244] proposing instead a framework where companies recognize their duty to protect and prioritize the interests of their users. This represents a significant departure from "surveillance capitalism," urging a reevaluation of business models that exploit personal information for profit.

Balkin's proposal that certain online services should be considered information fiduciaries who bear special duties of care, confidentiality, and loyalty towards users is crucially important when applied to GenAI systems. Like social media platforms, GenAI relies extensively on collecting and analyzing user data in order to function. Under an information fiduciary model, developers and providers of Generative AI would be obligated to act as fiduciaries, prioritizing user interests and welfare when handling their information. This marks a major shift from current incentives to exploit data for profit or capability gains. Instead, it emphasizes ethical standards of loyalty and care regarding user data and interactions.

This fiduciary approach takes on heightened importance given GenAI's ability to generate personalized content and recommendations based on analyzing a user's personal information and conversational patterns. The technology's capacity for mimicking users' individual speech habits underscores the need for their data to be handled responsibly under a fiduciary governance model. By placing at the center user protection and interests, rather than data exploitation, information fiduciary principles provide a means of fostering greater transparency and trust between GenAI systems and users. Applying these principles would promote human welfare over unchecked technological capability growth. Overall, Balkin's concept of information fiduciaries offers a good tool for policymakers to apply to the governance of Generative AI.

---

[240] *Id.* at 1209.
[241] *Id.* at 1208.
[242] *Id.* at 1209.
[243] *Id.* at 1226.
[244] Balkin, *To Regulate*, *supra* note 15, at 92.

### B.   *Platform Federalism and Reflecting Community*

We turn now to the second trust deficit facing digital media platforms: the fact that they are detached from any particular culture and locale, and therefore are necessarily unable to channel the traditional mechanisms of trust-building, deeply embedded within the cultural and societal fabric. The essence of trust, as rooted in shared values, communal goals, and cultural integration, presents a stark contrast to the global, culturally-detached nature of social media and GenAI.

Elsewhere, I have suggested integrating local, familiar institutions into the content moderation and curation processes of social media and search platforms, a concept that could be extended to GenAI.[245] This localization strategy aims to bridge the gap between global platforms and local cultural contexts, enhancing trust. This approach advocates for a structured involvement of domestic civil society in shaping online public dialogue, emphasizing the integration of local institutions like NGOs, media, and academia into the governance of digital platforms. It proposes that such inclusion can narrow the gap between global digital platforms and local communities, enhancing the relevance and responsiveness of online discourse.

By incorporating these local elements, digital media platforms can become more attuned to and reflective of the cultural and trust conditions of different communities. This approach could potentially connect these worldwide, detached platforms with local norms and values, addressing the challenges of establishing legitimacy in diverse cultural environments. It suggests legislative measures to ensure local civil society organizations play a significant role in content moderation, policy implementation, and establishing trusted information sources, aiming to reestablish their gatekeeping function in the digital age.

---

[245] Abiri & Guidi, *The Platform Federation*, *supra* note 70, at 5.

**GenAI and Content Moderation**: Formulating governance rules for GenAI, like content moderation policies for social platforms, requires profound local and cultural insight. It is up to civil society to imbue these principles with necessary nuance. This role, akin to civil society's proposed function in shaping content rules, exceeds mere oversight. It entails proactive engagement so GenAI follows an ethical framework that also resonates culturally. Moreover, platforms craft content policies through opaque processes, often in vague terms. While some standards prohibit certain expressions universally (e.g., blackface), most employ broad language compatible with diverse contexts. For instance, Facebook bans slurs that attack protected groups, but identifying slurs or their acceptable uses depends heavily on culture.[246] Thus, universal enforcement is impossible without applying specific social norms.

Accordingly, local civil society organizations should play a preeminent role in specifying how to implement these abstract standards. Local institutions are best suited to define acceptable speech bounds, humor contours, and satire limits for each jurisdiction. To enable civil society federalism, platforms must devise granular operational rules by region, with civil society input. Rather than platforms "training" civil society as "trusted flaggers," civil society should instruct platforms.

In the realm of GenAI, engagement of local civil society institutions in content moderation becomes essential for establishing local trust. Like human moderators on digital platforms, these local institutions should play a vital role in flagging and assessing content processed by GenAI. Crucially, this approach ensures that its algorithms stay informed by local civil society's understanding. From our viewpoint, a key goal of a trusted flagger system must be acknowledging and incorporating local speech norms into moderation. Since distilling these intricate norms into clear rules is impractical, achieving this necessitates direct involvement of local civil society in moderating. In summary, embedding local civil society institutions as core moderators of GenAI content can enable governance rooted in community norms and values.

**Model Training**: To build familiarity and trust, GenAI systems must become attuned to the cultural fabrics they operate within. This demands localization not just of policies and teams, but of the underlying algorithms themselves. Rather than monolithic models deployed indifferently worldwide, responsible GenAI requires an ensemble approach with diversity and specialization.

---

[246] *Facebook Community Standards: Hate Speech*, META TRANSPARENCY CENTER, https://transparency.meta.com/policies/community-standards/hate-speech/# [https://perma.cc/RJA6-BCNJ].

Developers could train core models on broad data, then refine regionally specific versions on localized examples. Knowledge bases could be populated with cultural background knowledge to ground reasoning. End-users could be able to provide context like country and language to adapt outputs. By learning cultural nuances, dialects, and norms, models can become simulacra embedded within each community.

Continuous retraining will update models on evolving locales. Testing localized iterations before launch will catch culturally aligned bugs. Partnerships with local researchers and civil society will imbue cultural wisdom. Hiring local teams and leaders will retain focus on community values. Advisory boards will guide alignment with norms.

In effect, GenAI models could have fluid personalities that shift appropriately across boundaries. They could speak with local tongues, argue with local logics, create with local aesthetics. Their synthetic eyes could recognize the world as a dynamic patchwork of cultures, seamlessly cross-stitching algorithms to suit each one.

### C.   *Digital Services Act vs. AI Act*

The discussion concludes by comparing the E.U.'s AI Act and Digital Services Act (DSA),[247] emphasizing the DSA's superior suitability for regulating GenAI's media dimensions, given its focus on online media platforms, making it more relevant than the U.S.'s AI Act for addressing the unique challenges posed by GenAI. This analysis supports the central premise of this Article, which argues that we should view GenAI not as a completely new and mysterious phenomenon of artificial intelligence, but rather as a continuation of the algorithmization of media. Therefore, applying ideas about the regulation of social media to GenAI allows us to more precisely pursue the goal of producing trustworthy intermediate information platforms.

The DSA introduces a multi-tiered framework of due diligence obligations designed to enhance the safety, transparency, and fairness of the digital ecosystem.[248] The DSA's purpose is to "reconcile the responsibilities of online platforms with their increased importance.[249]" It is therefore aimed exactly at digital media platforms. That said, it likely does not currently cover GenAI technologies, as these are tools for creating content rather than platforms disseminating third-party content.[250] The DSA targets entities that provide the infrastructure for hosting and sharing content across users, aiming to enhance moderation, transparency, and accountability.[251] Generative AI, in contrast, operates by generating new content from input data, likely positioning it outside the DSA's scope.[252] This distinction underscores the DSA's commitment to regulating the digital ecosystem's structural facets, rather than the content creation tools themselves. However, my purpose here is not to discuss actual legal application, but whether the DSA—which targets digital media harms—seems more appropriate to deal with the mediaaspects of GenAI than the AI Act.

---

[247] Regulation (EU) 2022/2065, 2022 O.J. (L 277) 1 [hereinafter DSA].
[248] *European Commission Policies, DSA: Making the Online World Safer*, Eur. Comm'n (Aug. 24, 2023), https://digital-strategy.ec.europa.eu/en/policies/safer-online [https://perma.cc/Y3RH-M3K5].
[249] Miriam C. Buiten, *The Digital Services Act: From Intermediary Liability to Platform Regulation*, 12 JIPITEC 361, 361 (2021).

At the foundational level, the DSA imposes universal obligations on all digital services eligible for liability exemptions.[253] This includes services like internet service providers, caching services, and web hosting services.[254] These basic obligations mandate the establishment of contact points for communication and the maintenance of transparency in how content moderation is conducted, ensuring accountability and accessibility in digital operations.[255]

Expanding upon the foundational requirements, the Digital Services Act specifies obligations for hosting services, emphasizing protocols for addressing illegal content and ensuring equitable moderation practices.[256] For online platforms, which include social networks, content-sharing services, and marketplaces, the Act introduces more detailed mandates.[257] These platforms are tasked with upholding higher standards in content moderation, designing services fairly, adhering to advertising protocols, and managing information amplification.[258] This tiered approach ensures that digital platforms facilitate safe and fair online environments for social engagement, commerce, and information sharing.

---

[250] Anthonia Ghalamkarizadeh, Telha Arshad, & Jasper Siems, *The Sorcerer's Apprentice Conundrum: Generative AI Content under the EU DSA and UK Online Safety Act*, Hogan Lovells: Engage (Jan. 24, 2024), https://www.engage.hoganlovells.com/knowledgeservices/news/the-sorcerers-apprentice-conundrum-generative-ai-content-under-the-eu-dsa-and-uk-online-safety-act [https://perma.cc/35BE-R8NX] (indicating that the DSA's language isn't a clear-cut fit when applied to GenAI use-cases); *see also* Philipp Hacker, *Generative AI at the Crossroads*, Oxford Bus. L. Blog (June 12, 2023), https://blogs.law.ox.ac.uk/oblb/blog-post/2023/06/generative-ai-crossroads [https://perma.cc/W892-LW98] ("The DSA does not apply to generative AI developers directly—this is a loophole that must urgently be fixed.").

[251] *The DSA Policy Essay*, *supra* note 30 ("The DSA regulates online intermediaries and platforms such as marketplaces, social networks, content-sharing platforms, app stores, and online travel and accommodation platforms.").

[252] Ghalamkarizadeh et al., *supra* note 250 (indicating that the DSA's language isn't a clear-cut fit when applied to GenAI use-cases).

[253] *The DSA Policy Essay*, *supra* note 30 ("All online intermediaries offering their services in the single market, whether they are established in the EU or outside, will have to comply with the new rules.").

[254] *Supra* note 247, at art. 3.

[255] *Id.*, art. 10.

[256] *Id.*, art. 6.

At the pinnacle of the DSA's regulatory structure are the special obligations designated for Very Large Online Platforms (VLOPs) and Very Large Online Search Engines (VLOSEs).[259] VLOPs are identified based on their extensive reach and impact, characterized by having a user base that represents a significant proportion of the EU's population.[260] This classification triggers the most stringent due diligence obligations, including comprehensive risk assessments, mitigation strategies, independent auditing, and crisis response mechanisms.[261] This tiered approach allows the DSA to scale its regulatory demands based on the potential impact and reach of digital services, ensuring a balanced yet effective governance model for the digital space.

When it comes to regulating GenAI as a far-reaching digital media platform, the E.U.'s Digital Services Act provides a more suitable framework than the narrower AI Act or Executive Order 14110.

First, the DSA bases oversight on platform size rather than risk categories.[262] This graduated approach is better adapted to media regulation, as scale correlates with societal impact. Larger platforms with expansive reach warrant more stringent supervision to maintain public trust. Proportional accountability also future-proofs regulations, allowing calibrated oversight as platforms grow.

---

[257] *Id.*, Ch. 3.; *see* Buiten, *supra* note 249, at 375.

[258] European Commission Policies, The Impact of the Digital Services Act on Digital Platforms, Eur. Comm'n (Nov. 3, 2023), https://digital-strategy.ec.europa.eu/en/policies/dsa-impact-platforms#:~:text=The%20DSA%20requires%20platforms%20to,cooperate%20with%20%E2%80%9Ctrusted%20flaggers%E2%80%9D [https://perma.cc/YU7M-LR2H].

[259] *See The DSA Policy Essay*, *supra* note 30; *see also* Buiten, *supra* note 249, at 367.

[260] DSA, *supra* note 254, art. 33.

[261] *See* Buiten, *supra* note 249, at 368.

[262] European Commission Press Release QANDA/20/2348, Questions and Answers: Digital Services Act (Dec. 19, 2023) ("With the Digital Services Act, unnecessary legal burdens due to different laws were lifted, fostering a better environment for innovation, growth and competitiveness, and facilitating the scaling up of smaller platforms, SMEs and start-ups.").

Second, core media functions necessitate transparency. Content moderation profoundly shapes online discourse yet remains opaque. The DSA mandates detailed disclosures and independent audits to surface how moderation systems operate. Scrutinizing these obscured but critical processes is crucial for oversight.

Third, visibility into curation and filtering algorithms that drive content recommendation and prioritization is imperative. The DSA requires transparency into the design and training data of such systems. This exposes any skewing of visibility and counters engagement-above-all optimization. Oversight of personalized advertising is also mandated.

Fourth, the DSA tackles the data exploitation characteristic of informational capitalism. It prohibits dark patterns that subvert consent and expands user data rights. Oversight of ad targeting algorithms is mandated to deter rights-violating microtargeting. This counters the surveillance advertising model.

Finally, large platforms must conduct annual assessments of potential societal harms. This holistic approach reaches beyond risk mitigation to align commercial incentives with democratic values. Proactive accountability discourages singular focus on profits over the public good.

This is not to suggest that the DSA offers an optimal solution for governing digital media platforms broadly. The DSA remains an imperfect work-in-progress. However, in contrast to the AI Act's narrow focus on technical risk management, the DSA holds significant advantages for regulating the uniquely media-centric functions and societal impacts of Generative AI models.

## Conclusion

As this Article illustrates, generative algorithms should not be viewed as some radical rupture necessitating unprecedented regulatory responses. Rather, situating GenAI within the trajectory of algorithmic mediation of the digital public sphere reveals it is the next phase of an ongoing process. Consequently, many strategies for governing GenAI can and should build upon existing and emerging models for regulating digital platforms like search engines and social media.

The path forward requires establishing GenAI systems as trusted intermediaries that foster a digital public sphere aligned with democratic values. This demands addressing the dual trust deficits stemming from misaligned incentives and the global-local divide. Beyond risk management, GenAI regulation must focus on reforms tailored to achieve this goal.

As the comparative analysis of the EU's AI Act and Digital Services Act illustrates, laws like the DSA designed explicitly to govern online platforms are better suited to regulate GenAI's societal impacts than general AI laws like the AI Act. Seeing GenAI as a continuation of the algorithmization of media and information highlights that existing conversations on platform governance must evolve to accommodate this new class of algorithmic intermediaries. But regulating GenAI does not require starting from scratch. Rather, it is the next chapter in an ongoing challenge - establishing trusted, democratically-aligned platforms to facilitate digital discourse.